\documentclass{article}
\usepackage[T1]{fontenc}

\usepackage{microtype}
\usepackage{graphicx}
\usepackage{subfigure}
\usepackage{booktabs}
\usepackage{hyperref}
\usepackage{xurl}
\usepackage{xcolor}
\usepackage[normalem]{ulem}
\usepackage{float}
\usepackage{enumitem}
\usepackage{array}
\usepackage{tabularx}
\usepackage{tikz}
\usetikzlibrary{arrows.meta,decorations.pathreplacing}
\usepackage{multicol}
\usepackage{fancyvrb}
\usepackage{pifont}
\usepackage{forest}
\useforestlibrary{edges}
\usepackage{etoc}
\usepackage{tcolorbox}
\tcbuselibrary{breakable,skins}
\tcbset{promptbox/.style={enhanced,breakable,colback=gray!4,colframe=gray!55,
  coltitle=black,colbacktitle=gray!18,fonttitle=\bfseries\small,
  fontupper=\small,boxrule=0.5pt,arc=1pt,left=8pt,right=8pt,top=6pt,
  bottom=6pt,toptitle=3pt,bottomtitle=3pt,boxsep=0pt,
  before upper={\setlength{\parskip}{3pt}}}}
\definecolor{figblue}{HTML}{2F6DB5}
\definecolor{figbluefill}{HTML}{E3EDF8}
\definecolor{figpanel}{HTML}{F4F5F7}
\definecolor{figpaneledge}{HTML}{7C838E}
\definecolor{hlcolor}{RGB}{255,236,179}
\newcommand{\sat}{\textcolor{green!45!black}{\ding{51}}}
\newcommand{\con}{\textcolor{red!75!black}{\ding{55}}}
\newcommand{\miss}{\textcolor{gray}{\ding{109}}}
\newcommand{\wt}[1]{\ifx&#1&\else[#1]\ \fi}
\newcommand{\nd}[3]{\wt{#1}\textbf{#2}\ \textcolor{gray}{#3}}
\newcommand{\cnd}[5]{\wt{#1}\textbf{#2}\ \textcolor{gray}{#3\ \ #4/#5}}
\newcommand{\lf}[4]{#1\ #2\ [#3] #4}
\newcommand{\dec}[1]{\textit{(#1)}}
\forestset{gangtree/.style={for tree={grow'=0,folder,folder indent=0.45em,
  font=\footnotesize,inner xsep=1.5pt,inner ysep=1.2pt,s sep=1pt,l sep=1.1em,
  edge={gray!60!black}}},
  hl/.style={fill=hlcolor,rounded corners=1pt}}

\let\origaddcontentsline\addcontentsline
\usepackage[accepted]{mlsys2025}
\makeatletter
\renewcommand{\Notice@String}{Preprint.}
\makeatother

\makeatletter
\g@addto@macro\@floatboxreset{\setlength{\parskip}{0pt}}
\def\@tablecaptype{table}
\long\def\@makecaption#1#2{
 \ifx\@captype\@tablecaptype\else\vskip 4pt\fi
        \baselineskip 11pt
        \setbox\@tempboxa\hbox{#1. #2}
        \ifdim \wd\@tempboxa >\hsize
        \sbox{\newcaptionbox}{\small\sl #1.~}
        \newcaptionboxwid=\wd\newcaptionbox
        \usebox\newcaptionbox {\footnotesize #2}
        \else
          \centerline{{\small\sl #1.} {\small #2}}
        \fi
 \ifx\@captype\@tablecaptype\vskip 4pt\fi}
\makeatother

\def\projectname{SLR-Eval}
\title{\projectname{}: Benchmarking LLM Agents on Real-World Software Specification Design}
\mlsystitlerunning{\projectname{}}

\begin{document}
\twocolumn[
\mlsystitle{\projectname{}: Benchmarking LLM Agents on Real-World Software Specification Design}
\begin{mlsysauthorlist}
\mlsyssetsymbol{equal}{*}
\mlsysauthor{Grant Hamblin}{equal,uoft}
\mlsysauthor{Kevin Song}{equal,uoft}
\mlsysauthor{Zhanda Zhu}{uoft}
\mlsysauthor{Anand Jayarajan}{uoft}
\mlsysauthor{Qidong Su}{nvidia,uoft}
\mlsysauthor{Sihang Liu}{uwaterloo}
\mlsysauthor{Nandita Vijaykumar}{uoft,vector}
\mlsysauthor{Gennady Pekhimenko}{uoft,vector,nvidia}
\end{mlsysauthorlist}
\mlsysaffiliation{uoft}{University of Toronto, Toronto, Ontario, Canada}
\mlsysaffiliation{uwaterloo}{University of Waterloo, Waterloo, Ontario, Canada}
\mlsysaffiliation{vector}{Vector Institute, Toronto, Ontario, Canada}
\mlsysaffiliation{nvidia}{NVIDIA, Toronto, Ontario, Canada}
\mlsyscorrespondingauthor{Grant Hamblin}{granthamblin0@gmail.com}
\mlsyscorrespondingauthor{Kevin Song}{xinyang.song@utoronto.ca}
\vskip 0.3in
\begin{abstract}
As software engineering (SWE) agents grow more proficient at coding, the bottleneck is shifting from implementation to designing specifications. However, existing coding benchmarks primarily evaluate \emph{implementation-level reasoning}: whether an agent can write code that satisfies a given specification. We argue that the next frontier for SWE agents is \emph{specification-level reasoning}: the ability to design system specifications that are well defined, consistent, and effective. To evaluate this capability, we introduce \projectname{}, a novel benchmark built from real Request for Comments (RFC) design proposals and maintainer reviews. \projectname{} comprises 58 tasks and 1,726 human-verified rubric criteria across seven production open-source systems (Linux, Kubernetes, Rust, Python, React, vLLM, and TVM). 
Given an initial specification, the codebase, and past RFCs, the agent must identify deficiencies in the proposed specification and design revisions to resolve them. Across 23 frontier SWE agents, the best-performing agent achieves a score of only 0.447 out of 1, and 35.6\% of expert-identified criteria are missed by every agent evaluated. We will open-source \projectname{} and its dataset construction pipeline.
\end{abstract}

]
\printAffiliationsAndNotice{\mlsysEqualContribution}

\suppressfloats[t]
\begin{figure}[t]
\small
\begin{minipage}{\columnwidth}
\emph{``The hardest single part of building a software system is deciding precisely what to build\ldots\ No other part of the work so cripples the resulting system if done wrong.''}\par
\raggedleft
--- \emph{The Mythical Man-Month}~\citep{brooks1995mythical}\par
\end{minipage}
\end{figure}

\begin{figure*}[t]
\centering
\includegraphics[width=\textwidth]{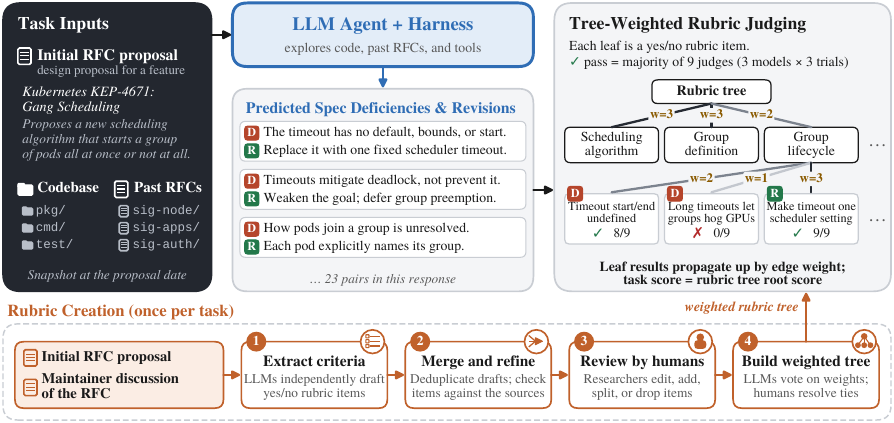}
\caption{\projectname{} on one task, Kubernetes KEP-4671 (gang scheduling). \emph{Top:} Given an initial RFC (design proposal), the codebase, and past RFCs as of the proposal date, an agent predicts specification deficiencies (D) and revisions (R); nine LLM judges score them against a tree-weighted rubric. \emph{Bottom:} The rubric is built from the expert review that followed the proposal, which the agent never sees.}
\label{fig:overview}
\end{figure*}

\section{Introduction}

The capabilities of software engineering (SWE) LLM agents have increased rapidly~\citep{yang2024sweagent,wang2025openhands,anthropic2025claudecode,openai2025codex}. Since 2024, agents have progressed from resolving isolated issues to taking on long-horizon engineering work that requires hours of autonomous effort, such as implementing features across large repositories and rebuilding entire libraries~\citep{huang2026deepswe,desai2026swemarathon}. On SWE-bench Verified~\citep{jimenez2024swebench,chowdhury2024swebenchverified}, a standard benchmark where agents resolve GitHub issues against repository test suites, the top resolution rate rose from 4.4\% to 79.2\% by December 2025~\citep{swebench2026leaderboard} and 97\% by September 2026~\citep{vals2026swebench}.

As agents become more capable at coding, the bottleneck in software engineering shifts from implementing code to designing system specifications: reasoning about a system's behavior, interfaces, architecture, and design tradeoffs~\citep{brooks1995mythical,stoica2024specifications}. For example, in a study of 132 Anthropic engineers and researchers, participants reported the smallest AI productivity gains in design/planning, whereas implementation/debugging showed the largest gains~\citep{anthropic2025work}. Similarly, a lead .NET runtime Microsoft engineer reports that as agents wrote more of the code, work shifted from implementation to ideation, design, and architecture~\citep{toub2026copilot}. 

We argue that the next generation of SWE agents must be able to design sound system specifications, one of the most challenging parts of software engineering~\citep{brooks1995mythical,stoica2024specifications}. First, specification flaws are pervasive in practice: practitioners at 109 of 228 surveyed companies ranked incomplete or hidden requirements among their five most important requirements-engineering problems~\citep{mendezfernandez2017naming}. Furthermore, a flawed specification has large downstream effects: in one study of network-partition failures in distributed systems, failures caused by design flaws took a mean of 205 days to resolve, compared with 81 days for failures caused by implementation flaws~\citep{alquraan2018partitions}.

We observe that existing coding benchmarks have not adapted to this shift in bottleneck from implementation to specification. Across all SWE benchmarks used by 39 flagship LLM releases from October 2025 to October 2026 (Appendix~\ref{app:benchmark-survey}), 24 out of 26 (92.3\%) benchmarks operate under the same assumption: the specification is given to the agent as correct and complete, and the benchmark grades only whether the agent's code behaves as specified.\footnote{The other two check code against its intended behavior, not whether that behavior is sound. See Section~\ref{sec:related-work}.}
While recent benchmarks continues to challenge SWE agent via longer horizons~\citep{desai2026swemarathon}, terminal environments~\citep{merrill2026terminalbench}, and whole-program reconstruction~\citep{yang2026programbench}, no major benchmark evaluates whether the agent can design the specification itself.

\begin{figure}[t]
\centering
\begin{tikzpicture}[
    stage/.style={draw=figblue, fill=figbluefill, rounded corners=2pt, minimum width=1.6cm, minimum height=1.4cm},
    steplabel/.style={align=center, text width=1.5cm, font=\footnotesize, inner sep=0pt},
    flow/.style={-{Latex[length=1.6mm]}, thick},
    pics/doc/.style={code={
        \draw[fill=white] (-0.15,-0.21) -- (-0.15,0.21) -- (0.07,0.21) -- (0.15,0.13) -- (0.15,-0.21) -- cycle;
        \draw (0.07,0.21) -- (0.07,0.13) -- (0.15,0.13);
        \foreach \y in {0.06,-0.03,-0.12} {\draw (-0.09,\y) -- (0.09,\y);}
    }},
    pics/person/.style={code={
        \fill (0,0.02) circle (0.07);
        \fill (-0.12,-0.21) arc[start angle=180, end angle=0, radius=0.12] -- cycle;
    }},
    pics/bubble/.style={code={
        \draw[fill=white, rounded corners=2pt] (-0.14,-0.07) rectangle (0.14,0.07);
        \draw[fill=white] (-0.05,-0.07) -- (-0.09,-0.14) -- (0.0,-0.07);
        \foreach \x in {-0.06,0,0.06} {\fill (\x,0) circle (0.015);}
    }},
    pics/check/.style={code={
        \fill (0,0) circle (0.09);
        \draw[white, line width=0.9pt] (-0.045,0.0) -- (-0.01,-0.035) -- (0.045,0.035);
    }},
    pics/terminal/.style={code={
        \fill[rounded corners=1.5pt] (-0.27,-0.2) rectangle (0.27,0.2);
        \fill[black!50] (-0.27,0.1) rectangle (0.27,0.15);
        \node[white, font=\ttfamily\bfseries\scriptsize, anchor=west, inner sep=0pt] at (-0.21,-0.03) {>\_};
    }}
]
    \node[stage] (proposal) at (0,0) {};
    \node[stage] (review) at (1.86,0) {};
    \node[stage] (spec) at (3.72,0) {};
    \node[stage, draw=figpaneledge, fill=figpanel, minimum width=2.2cm] (code) at (5.84,0) {};
    \pic at (0,0.36) {doc};
    \pic at (1.76,0.32) {person};
    \pic at (1.98,0.32) {person};
    \pic at (2.08,0.54) {bubble};
    \pic at (3.72,0.36) {doc};
    \pic at (3.86,0.21) {check};
    \pic at (5.84,0.34) {terminal};
    \node[steplabel] at (0,-0.24) {Initial\\design};
    \node[steplabel] at (1.86,-0.24) {Expert\\review};
    \node[steplabel] at (3.72,-0.24) {Accepted\\spec};
    \node[steplabel, text width=2.1cm] at (5.84,-0.24) {Implement\\and test};
    \draw[flow] (proposal) -- (review);
    \draw[flow] (review) -- (spec);
    \draw[flow] (spec) -- (code);
    \draw[decorate, decoration={brace, mirror, amplitude=4pt}] (-0.8,-0.85) -- (4.52,-0.85);
    \node[align=center, font=\footnotesize, anchor=north] at (1.86,-1.0) {\textbf{Specification-}\\\textbf{level reasoning}\\(\projectname{})};
    \draw[decorate, decoration={brace, mirror, amplitude=4pt}] (4.74,-0.85) -- (6.94,-0.85);
    \node[align=center, font=\footnotesize, anchor=north, inner xsep=0pt] at (5.84,-1.0) {\textbf{Implementation-}\\\textbf{level reasoning}\\(SWE-bench and others)};
\end{tikzpicture}
\caption{Specification-level vs. implementation-level reasoning.}
\label{fig:development-lifecycle}
\end{figure}

\begin{table}[t]
\centering\footnotesize
\setlength{\tabcolsep}{3.5pt}
\caption{Implementation-level and specification-level reasoning, as evaluated by SWE-bench and \projectname{}.}
\label{tab:levels}
\begin{tabularx}{\columnwidth}{@{}>{\raggedright\arraybackslash}p{27pt}>{\raggedright\arraybackslash}p{84pt}>{\raggedright\arraybackslash}X@{}}
\toprule
 & \textbf{Implementation-level}\newline\textbf{reasoning} & \textbf{Specification-level}\newline\textbf{reasoning} \\
\midrule
Focus & Does the code\newline meet the spec? & Is the spec sound for the system and its goals? \\
\addlinespace
Input & Spec, codebase & Proposed spec, codebase,\newline design history \\
\addlinespace
Output & Code changes & Deficiencies and revisions \\
\addlinespace
\addlinespace
Skills & Read code;\newline program, test, debug & Understand system design;\newline anticipate consequences;\newline weigh tradeoffs \\
\bottomrule
\end{tabularx}
\end{table}

As illustrated in Figure~\ref{fig:development-lifecycle}, existing benchmarks such as SWE-bench focus on the final stage, which we refer to as \emph{implementation-level reasoning}: the ability to realize a given specification in code. In contrast, we define \emph{specification-level reasoning} as the ability to develop specifications that state the required behavior completely and unambiguously (\emph{well defined}), align with the existing system's interfaces and design philosophy (\emph{consistent}), and solve the motivating problem (\emph{effective})~\citep{iso2018requirements}. Table~\ref{tab:levels} compares the two types of reasoning. Importantly, the latter requires understanding system-wide architecture and past design decisions, anticipating the consequences of a design choice, and weighing competing tradeoffs.

However, evaluating specification-level reasoning poses a key challenge: obtaining real-world tasks and ground truth for whether a specification is sound. The data sources used by prior coding benchmarks, such as GitHub issues, pull requests, and test suites, cannot be used here because they take the specification as given. \textbf{Our key insight is to utilize the Request for Comments (RFC) process as a data source} for constructing realistic and challenging benchmark tasks~\citep{kubernetes2026kep,rust2026rfcs,warsaw2000pep1,linux2026submitting}. In the RFC process, before implementing a major feature or architectural change, a contributor submits a structured proposal that details the motivation, proposed mechanisms, and design tradeoffs. The project's maintainers and domain experts then review and iterate on the proposal in public until they reach an accepted specification.\footnote{Example: \href{https://github.com/kubernetes/enhancements/blob/master/keps/sig-scheduling/4671-gang-scheduling/README.md}{Kubernetes gang scheduling design document} and \href{https://github.com/kubernetes/enhancements/pull/5558}{expert review discussion}.}
This RFC review process is the quality gate used to decide whether a design is ready to be implemented. The RFC proposal is reviewed by core project contributors (each reviewer behind \projectname{}'s data source has an average of 4.3 years of tenure and 559 prior commits to the project's main repository), and each RFC discussion consists of an average of 94 comments from 15 reviewers. We therefore use the collective judgment recorded in these expert RFC reviews as the ground-truth reference for our benchmark.

In each RFC discussion, experts identify problems in the initial proposal (\emph{deficiencies}) and agree on changes to resolve them (\emph{revisions}). We turn each RFC into a benchmark task by extracting the deficiencies and revisions identified by the human experts into a grading rubric, and reconstructing the project's environment as of the initial proposal date. During evaluation, the agent receives the initial design proposal, the codebase, and past RFCs, and predicts deficiencies and revisions, which we grade against the rubric (Figure~\ref{fig:overview}).
In addition, constructing a benchmark from RFC discussions introduces two grading challenges. \textbf{(1) Grading open-ended predictions:} Since we utilize historical expert discussions as source of truth, agent predictions that match this ground truth are verifiably valid, but unmatched predictions could still be valid critiques of the specification that the experts did not raise. To address this, we adopt the evaluation methodology from information retrieval~\citep{zobel1998reliable,buckley2004incomplete}. Since we cannot reliably grade predictions outside of the ground truth, we leave unmatched predictions as \emph{unjudged} (neither credit nor penalty). Predictions that directly match ground truth receive credit. To prevent reward hacking through excessive guessing, we penalize predictions that directly contradict the ground truth, similar to HealthBench~\citep{arora2025healthbench}.
\textbf{(2) Weighting criteria by importance:} Specification deficiencies and revisions vary in importance, so an equally weighted grading rubric would over-reward minor issues at the expense of critical flaws. For instance, with a new scheduling rule, ensuring that the scheduler is deadlock-free should weigh more than specifying which observability metrics to expose. To reflect this, we represent each task's rubric as a hierarchical weighted tree grouped by topic. During evaluation, a task's total score is calculated from the relative weights of the criteria that the agent satisfies (minus any contradiction penalties), where a score of 1.0 corresponds to fully matching the collective judgment of expert human maintainers.

Based on these insights, we present \projectname{}, a novel specification-level reasoning benchmark for SWE agents. \projectname{} comprises 58 tasks and 1,726 human-verified rubric criteria from seven open-source projects spanning the systems stack: an OS kernel (Linux), cluster orchestration (Kubernetes~\citep{burns2016borg}), programming languages (Rust and Python), a UI framework (React), and ML systems (vLLM~\citep{kwon2023vllm} and TVM~\citep{chen2018tvm}).

Using \projectname{}, we evaluate 23 frontier SWE agents and observe that current agents struggle with specification-level reasoning: the highest-scoring agent (Claude Opus 5.5 at max effort) achieves a score of 0.447 out of 1.0. Moreover, 35.6\% of the rubric criteria are not satisfied by any agent we tested. To better understand where agents succeed and fail, we define a taxonomy of specification deficiencies (Figure~\ref{fig:category-performance}). Agents perform worst on the \emph{scope/rationale/high-level design decisions} category (11.1\% average score), even though it is one of the most common types of deficiencies raised by human experts (14.0\%, third largest of 12 categories). Our results support practitioners' observations that current agents struggle with high-level design decisions, as discussed earlier~\citep{toub2026copilot,anthropic2025work}. We hope \projectname{} will help establish a rigorous foundation for measuring and advancing specification-level reasoning in the next generation of SWE agents.

In summary, we make the following contributions:
\begin{itemize}[nosep,leftmargin=*]
    \item We formulate \emph{specification-level reasoning}, identifying a critical capability bottleneck for SWE agents that prior benchmarks overlook.
    \item We design an evaluation methodology that turns open-source Request for Comments (RFC) discussions into specification-level reasoning tasks, using IR-based unjudged scoring and hierarchical weighted rubric trees to grade open-ended design predictions.
    \item We build \projectname{}, a benchmark of 58 tasks across seven production open-source projects with 1,726 human-verified rubric criteria. 
    \item We evaluate 23 state-of-the-art SWE agents to collect 1,334 agent trajectories and analyze their performance, cost, and failure patterns. Even the strongest agent achieves a score of only 0.447 out of 1.0, revealing a large gap between current SWE agents and human experts on specification-level reasoning.
\end{itemize}

\section{Background and Motivation}

\subsection{SWE Agents and Their Benchmarks}
\label{sec:swe-benchmarks}
SWE agents are LLM-based systems that autonomously perform software engineering tasks by reading code, editing files, and executing commands within a software environment~\citep{yang2024sweagent,wang2025openhands,anthropic2025claudecode,openai2025codex}. SWE-bench established the standard methodology for evaluating SWE agents~\citep{jimenez2024swebench}: in each task, the agent under test is given a codebase and a description of the desired change, and modifies the codebase to implement that change. The code edits are then graded using a predefined set of tests. SWE-bench constructs these tasks by mining real GitHub issues, pull requests, and human verified test cases.

The capabilities of SWE agents have improved rapidly. On SWE-bench Verified~\citep{chowdhury2024swebenchverified}, agent resolution rates rose from 4.4\%~\citep{jimenez2024swebench} to 79.2\% by December 2025~\citep{swebench2026leaderboard} and 97\% by September 2026~\citep{vals2026swebench}, effectively saturating the benchmark~\citep{liu2026converged}. To create more challenging SWE tasks, subsequent benchmarks have broadened the scope of work to long-horizon tasks~\citep{huang2026deepswe}, whole-program reconstruction~\citep{yang2026programbench}, and terminal environments~\citep{merrill2026terminalbench}.
Yet they retain the same underlying evaluation paradigm: the benchmark assumes a sound specification is given, and grades only whether the agent's code satisfies it. We refer to this capability as \emph{implementation-level reasoning}: the ability to implement a given specification of desired behavior in code.

\subsection{Specification-Level Reasoning}
\label{sec:spec-level-reasoning}

We argue that evaluating only implementation-level reasoning is insufficient for the next generation of SWE agents. First, designing system specifications is known to be one of the most challenging parts of software engineering~\citep{brooks1995mythical,stoica2024specifications}. In practice, specification defects are a pervasive problem due to hidden assumptions, omissions, and design flaws~\citep{mendezfernandez2017naming}. Second, since a specification governs behavior and interfaces across components, such flaws have large downstream consequences~\citep{alquraan2018partitions}. Finally, as agents automate code implementation, the bottleneck to achieving truly autonomous software engineering shifts to reasoning about and designing high-level specifications~\citep{toub2026copilot,anthropic2025work}.

To capture this capability, we define \textbf{specification-level reasoning} as the ability to develop specifications that are well defined, consistent, and effective. Drawing on the ISO/IEC/IEEE 29148 standard~\citep{iso2018requirements}, a specification is \emph{well defined} if it states the required behavior and constraints unambiguously. It is \emph{consistent} if it conflicts neither with the existing system nor the project's design philosophy (e.g., the Rust community's strictness on memory safety). Finally, it is \emph{effective} if the system behavior solves the problem the motivated the change.

Specification-level reasoning requires fundamentally different capabilities than implementation-level reasoning (Table~\ref{tab:levels}). Whereas the latter tests an agent's ability to comprehend, develop, test, and debug code against a given specification, the former requires understanding system design philosophy, anticipating the consequences of design choices, and weighing various tradeoffs. Despite its importance, prior coding benchmarks have overlooked this capability. In fact, SWE-bench Verified explicitly filter out tasks with underspecified issue descriptions, treating specification ambiguity as a benchmark defect~\citep{chowdhury2024swebenchverified}.

\subsection{Example: Gang Scheduling in Kubernetes}
\label{sec:gang-example}

We illustrate the difference between implementation-level and specification-level reasoning with a running example: gang scheduling in Kubernetes. In Kubernetes, an application runs as one or more \emph{pods}: logical units of containerized processes that request hardware resources such as CPUs and GPUs. By default, the Kubernetes scheduler assigns each pod to physical cluster nodes independently. However, tightly coupled parallel applications, such as distributed AI training, run as a \emph{pod group} that makes progress only when all of its pods are running simultaneously. Scheduling such groups one pod at a time can cause resource deadlock: suppose two pod groups each need four pods, each pod needs one GPU, and the cluster has four GPUs. If the default scheduler assigns two pods from each group, each group waits for GPUs the other holds, and neither proceeds (Appendix Figure~\ref{fig:gang-scheduling-scenario}). \emph{Gang scheduling}~\citep{feitelson1992gang} solves this problem by scheduling a pod group atomically, allocating resources to all pods in the group together or none of them.

The initial proposal for supporting gang scheduling in the default Kubernetes scheduler (Appendix Figure~\ref{fig:gang-initial-rfc}) makes three explicit design decisions:
\begin{enumerate}[nosep,leftmargin=*]
    \item \emph{Group definition.} An application can divide its pods into multiple pod groups, each scheduled as a separate gang.
    \item \emph{Group placement.} Rather than allocating resources to a whole group at once, the scheduler still assigns pods to nodes one at a time, holding each pod's allocated resources until every pod in the group is assigned.
    \item \emph{Waiting timeout.} Each application configures a timeout: if its group cannot acquire resources for all of its pods before the timeout expires, the scheduler releases the resources already allocated to the group.
\end{enumerate}

An implementation-level benchmark would give an agent these decisions as its specification and grade whether the agent implements them correctly. However, as surfaced during the Kubernetes design review (Appendix Table~\ref{tab:gang-review}), this initial specification is flawed:
\begin{enumerate}[nosep,leftmargin=*]
    \item \emph{Group definition: not well defined.} With multiple pod groups, the proposal does not specify whether one group is allowed to start running if another group in the same application cannot be scheduled.
    \item \emph{Group placement: ineffective.} Because the scheduler still assigns pods one at a time while holding their allocated resources, two groups can still split the cluster's GPUs and deadlock, failing to solve the motivating problem.
    \item \emph{Waiting timeout: inconsistent.} In Kubernetes's design philosophy, cluster-wide resource sharing is controlled by administrators rather than individual applications. Letting each application set its own timeout conflicts with this principle, as an application can set a long timeout to hold scarce GPUs and block other jobs.
\end{enumerate}

Catching and fixing such flaws at the specification stage is critical given the scale of downstream implementation: within a year of the proposal, at least 23 merged pull requests implemented, tested, or refined gang scheduling and its supporting scheduler interfaces, totaling about 32,800 lines of changes, including generated API code and tests.

\section{\projectname{} Design}
In this section, we summarize the key challenges and how
\projectname{} addresses them.

\subsection{A Data Source for Specification-Level Reasoning}
\label{sec:challenge-data}

\textbf{Challenge.} To evaluate specification-level reasoning, a benchmark needs real design proposals for agents to analyze and an authoritative reference to grade agent predictions against. Prior benchmarks such as SWE-bench build tasks from real Github issue and pull requests, and grade the agent predictions with human written tests. However, such data sources cannot evaluate specification-level reasoning because they take the specification as given.

\textbf{Solution.} Our key insight is to build \projectname{} from the Request for Comments (RFC) process used by mature open-source projects. RFC is a formal process used to propose, debate, and document major architectural decisions \textit{before} implementation~\citep{kubernetes2026kep,rust2026rfcs,warsaw2000pep1,linux2026submitting,react2026rfcs,vllm2026contributing,tvm2026rfcs}.\footnote{Different projects may use various names to refer to RFCs, e.g., KEPs for Kubernetes, PEPs for Python.} Initially, an author submits a design proposal describing a specific change~\citep{rust2026rfcs,kubernetes2026kep}, which specifies the motivation, proposed mechanisms, and design alternatives. Maintainers and domain experts then conduct a multi-round public review as the author revises the document toward acceptance or rejection. The initial and final versions of the design documents, along with the record of expert critiques are archived within this process.

Building \projectname{} from RFCs offers several benefits:
\begin{itemize}[nosep,leftmargin=*]
    \item \textbf{Real-world complexity.} RFCs capture consequential architectural decisions in widely used production systems (e.g., vLLM, Linux, and Kubernetes) that span multiple components and have broad downstream impact.
    \item \textbf{Explicit reasoning traces.} Rather than recording only the final outcome, RFC discussions make the step-by-step specification level reasoning visible, documenting the deficiencies raised, the design trade-offs debated, and how reviewers arrived at the final revisions.
    \item \textbf{Expert authority.} Reviews are conducted by the project's core contributors, who have the deepest domain expertise with each system (each reviewer behind \projectname{}'s data source has an average of 4.3 years of tenure and 559 prior commits to the project's main repository).
    \item \textbf{Continual supply.} Active open-source projects continuously propose and review new RFCs as their systems evolve, providing an ongoing stream of tasks to expand the benchmark over time.
\end{itemize}

To turn each RFC into a benchmark task, we extract the initial proposal and the expert discussion that followed it to build a ground-truth rubric (Section~\ref{sec:impl-rubrics}). The rubric captures two types of criteria: \emph{deficiencies} (problems in the initial specification) and \emph{revisions} (changes to the specification to address deficiencies). During evaluation, the agent is given the initial design proposal, a snapshot of the project's codebase, and past RFCs up to the proposal's submission date (Figure~\ref{fig:overview}; Section~\ref{sec:impl-environments}). The agent predicts a list of deficiencies and revisions targeting the initial specification, which we grade against the ground-truth rubric.

\subsection{Grading Open-Ended Predictions}
\label{sec:challenge-grading}

\textbf{Challenge.} Grading agent predictions in \projectname{} is asymmetric. \projectname{} takes the final design accepted by human experts as ground truth, so agent prediction that matches the rubric is verifiably valid. However, predictions that do not match the rubric are unverifiable, as they could represent valid critiques or alternative designs that human experts did not consider. Thus, penalizing all unmatched predictions can unfairly punish valid insights from the agent.

\textbf{Solution.} Inspired by information retrieval (IR) evaluations under incomplete relevance judgments~\citep{manning2008ir,zobel1998reliable,buckley2004incomplete,sakai2007bpref}, we treat predictions that match no rubric criterion as \emph{unjudged}, receiving neither credit nor penalty. Agent predictions that directly match a rubric criterion receives credit.  To prevent agents from reward-hacking by generating a large number of predictions to maximize credits, we penalize predictions that directly contradicts a rubric criterion, similar to negative-point rubrics in HealthBench~\citep{arora2025healthbench}.

\subsection{Weighting Criteria by Importance}
\label{sec:challenge-weighting}

\textbf{Challenge.} In real-world specification-level reasoning, not all deficiencies and revisions are equally important. For example, in Kubernetes gang scheduling, ensuring that the scheduler is deadlock-free matters much more than specifying which observability metrics to expose. A grading rubric that weights every criterion equally would over-reward minor issues at the expense of critical flaws.

\textbf{Solution.} To reflect this relative importance, we represent each task's rubric as a hierarchical weighted tree grouped by topic, building upon PaperBench~\citep{starace2025paperbench} (Figure~\ref{fig:overview}; Section~\ref{sec:impl-rubrics}). We place each rubric criterion as a leaf node in the tree, and group related leaf nodes by topic, e.g., safety, compatibility, performance (limiting maximum tree depth to three). We then assign each node (both leaf and internal nodes) a weight of 1, 2, or 3 indicating its importance relative to its siblings. During evaluation, each leaf node receives $1$ if satisfied, $-1$ if contradicted, and $0$ otherwise (Section~\ref{sec:impl-judging}). Each internal node's score is computed as the weighted average of its children's scores, propagating up to the root to produce the final task score, where a score of 1.0 corresponds to fully matching the collective judgment of expert human maintainers. In Figure~\ref{fig:overview}, for example, the group lifecycle node has three leaf nodes with weights 2, 1, and 3; satisfying the first and third leaves yields a score of $(2 \cdot 1 + 1 \cdot 0 + 3 \cdot 1)/6 = 0.833$.

\section{Benchmark  Implementation}

\subsection{Task Collection and Rubric Construction}
\label{sec:impl-rubrics}

\projectname{} contains 58 tasks from seven projects: ten from TVM and eight each from Kubernetes, vLLM, React, Rust, Linux, and Python (Appendix~\ref{app:tasks}). We select proposals that maintainers eventually accepted after substantial expert discussion. We build each task's rubric in four steps (Figure~\ref{fig:overview}, bottom; see Appendix~\ref{app:gang-rubric} for an example rubric):

\textbf{Extract criteria.} Three LLM agents from distinct model families (GPT, Claude, and Gemini) each receive the initial proposal and the historical expert discussion, and independently extract a flat list of candidate rubric criteria (Table~\ref{tab:rubric-pipeline}). Using multiple independent agents increases coverage of issues raised across long discussion threads. Each criterion is written as a self-contained yes/no rubric criterion.

\textbf{Merge and refine.} An LLM consolidates the three candidate lists by meaning, preserving distinct discussion-supported criteria, removing duplicates, and resolving conflicts against the source material. Next, a quality-review pass scores each merged criterion against the source documents across ten rubric-quality dimensions (e.g., source grounding, atomicity, and binary judgeability; Appendix~\ref{app:rubric-extraction}) and recommends whether to keep, revise, split, merge, or drop the item. A separate revision pass then verifies each recommendation against the initial proposal and expert discussion before applying the edits.

\textbf{Human review.} Graduate-level researchers review all 1,709 refined criteria, and each criterion receives two independent reviews against the initial proposal and expert discussion to verify that every item accurately reflects the experts' deliberation (Appendix~\ref{app:human-review}). Because the expert discussion already establishes the technical judgments, researchers verify grounding and refine phrasing rather than authoring domain critiques from scratch. In total, the researchers modified or removed 412 criteria (24\%: 195 revised, 96 split, 46 merged, and 75 dropped), producing 1,726 finalized criteria.

\textbf{Weighted tree.} An LLM organizes the finalized criteria by topic into a hierarchy of at most four levels without altering the human-verified criterion text, placing each criterion as a leaf node (Section~\ref{sec:challenge-weighting}; Appendix~\ref{app:tree-weighting}). Given the complete tree, initial proposal, and expert discussion, three LLM judges independently compare sibling nodes from the bottom up and assign each non-root node a relative importance weight of $1$, $2$, or $3$ with a written rationale. Deterministic majority voting merges the three weight assignments, and researchers inspect all node weights and resolve any ties to produce the final rubric.

\subsection{Agent Evaluation Environment}
\label{sec:impl-environments}

To evaluate an agent on a task, we place it in an isolated, offline workspace that reconstructs the project's state when the initial proposal was submitted (Appendix~\ref{app:environments}). To prevent temporal leakage, we check out the codebase and past RFCs at the last commit before the submission date, delete all later commits, and provide an offline tool to search prior RFC discussions created before that date (excluding the target RFC). Given the initial proposal, the agent may freely explore these resources and write or run code, with no limits on time, tokens, or prediction count (Appendix~\ref{app:agent-prompt}).

\subsection{Prediction Judging and Score Aggregation}
\label{sec:impl-judging}

To grade an agent's predictions, we use an ensemble of three LLM judges~\citep{zheng2023judging,verga2024juries} (GPT-6.1 Sol, Claude Opus 5.5, and Gemini 3.8 Flash; Appendix~\ref{app:judge-prompt}). Each judge evaluates every rubric criterion three times against the agent's predictions, producing nine verdicts per criterion. We map each verdict to $+1$ (satisfied), $-1$ (contradicted), or $0$ (unmatched) and take the median of the nine verdicts as the leaf score. Finally, we propagate the leaf scores up the weighted rubric tree to compute the task score (Section~\ref{sec:challenge-weighting}; Appendix~\ref{app:score-definition}), and average across all 58 tasks to obtain the final benchmark score.

\section{Evaluation}
\label{sec:evaluation}

We evaluate 23 agents on \projectname{}, examining overall performance and cost (\S\ref{sec:main-results}), where agents succeed and fail (\S\ref{sec:where-fail}), agent behavior and dynamics (\S\ref{sec:what-separates}), and scoring reliability (\S\ref{sec:judge-reliability}). Table~\ref{tab:eval-setup} summarizes the evaluation setup, and Appendix Table~\ref{tab:eval-models} reports the full numerical results.

\begin{figure*}[t]
    \centering
    \includegraphics[width=\textwidth]{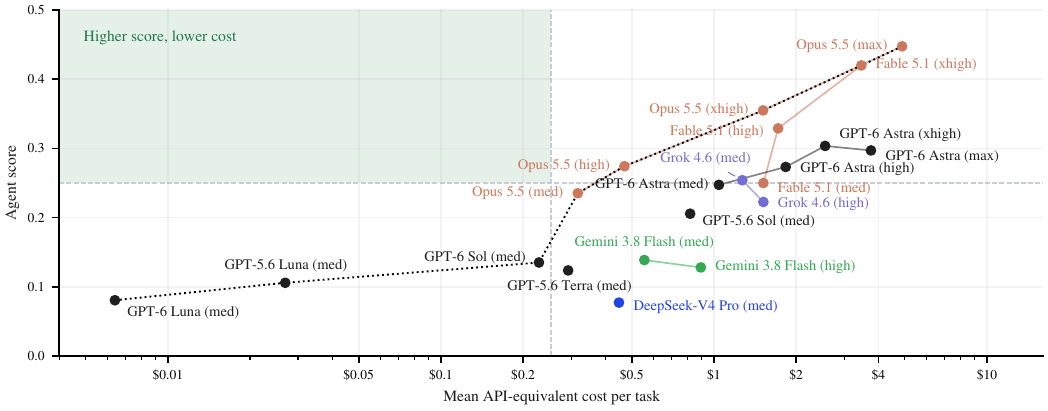}
    \caption{Score versus cost per task (log scale). Dotted line shows Pareto frontier. Solid lines connect effort settings of the same model. }
    \label{fig:eval-frontiers}
\end{figure*}

\begin{table}[t]
    \centering\footnotesize
    \setlength{\tabcolsep}{4pt}
    \caption{Evaluation setup.}
    \label{tab:eval-setup}
    \begin{tabularx}{\columnwidth}{@{}>{\raggedright\arraybackslash}p{38pt}>{\raggedright\arraybackslash}X@{}}
        \toprule
        Tasks & 58 proposals from 7 open-source projects \\
        \addlinespace
        Agents & Native agent harnesses for GPT (Codex), Claude (Claude Code), Gemini (Antigravity), and Grok (Grok Build). For open-source models such as DeepSeek, we use the Codex harness \\
        \addlinespace
        Judging & Each prediction scored by GPT-6.1 Sol, Claude Opus 5.5, and Gemini 3.8 Flash, 3 trials each (9 total)\\
        \addlinespace
        Scoring & Per-task score: tree-weighted rubric. Per-agent score: mean of all task scores \\
        \addlinespace
        Cost & API-equivalent dollars from recorded token usage at providers' standard rates (October 2026) \\
        \bottomrule
    \end{tabularx}
\end{table}

\subsection{Overall Agent Performance and Cost}
\label{sec:main-results}

We first evaluate agent performance on \projectname{} across projects and inference costs.

\textbf{Even the strongest agent scores less than 50\%.} Figure~\ref{fig:eval-frontiers} plots each agent's score against its mean API cost per task. Claude Opus 5.5 (max) achieves the highest score at 0.447 out of 1.0, followed by Fable 5.1 (xhigh) at 0.420. Across the entire benchmark, 35.6\% of the 1,726 rubric criteria are not satisfied by any agent we evaluated. This reveals a large capability gap between current SWE agents and human experts, leaving substantial headroom for future agents.

\begin{figure}[t]
    \centering
    \includegraphics[width=\columnwidth]{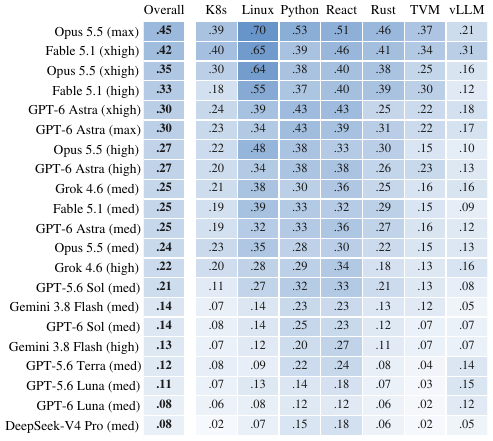}
    \caption{Overall and per-domain scores breakdown.}
    \label{fig:eval-breakdowns}
\end{figure}

\textbf{The capability gap holds across all projects.} Figure~\ref{fig:eval-breakdowns} breaks scores down by project. Averaged across all agents, scores are lowest on vLLM (0.13) and TVM (0.15), and highest on Linux (0.31) and React (0.31). Even on the highest-scoring projects, however, the strongest agent remains far below expert review.

\textbf{Higher cost improves scores on the Pareto frontier, but higher reasoning effort does not improve every model.} Eight agents form the cost--performance Pareto frontier (Figure~\ref{fig:eval-frontiers}). Across all agents, score correlates strongly with API cost (Spearman $\rho=0.80$). Mid-tier effort settings offer a favorable cost--performance tradeoff: Opus 5.5 (high) scores 0.274 at \$0.47 per task, retaining 61\% of the top score at 10\% of the cost. Across effort settings, Opus 5.5 and Fable 5.1 improve substantially with higher reasoning effort. 
By contrast, GPT-6 Astra (max), Gemini 3.8 Flash, and Grok 4.6 score slightly lower at higher effort settings. Contradictions cause much of this drop: without the contradiction penalty, Astra and Gemini scores stay flat, and the Grok score drops half as much.

\subsection{Where Agents Succeed and Fail}
\label{sec:where-fail}

Next, we analyze why agents struggle and which types of deficiencies and revisions they fail on most often.

\begin{figure}[t]
    \centering
    \includegraphics[width=\columnwidth]{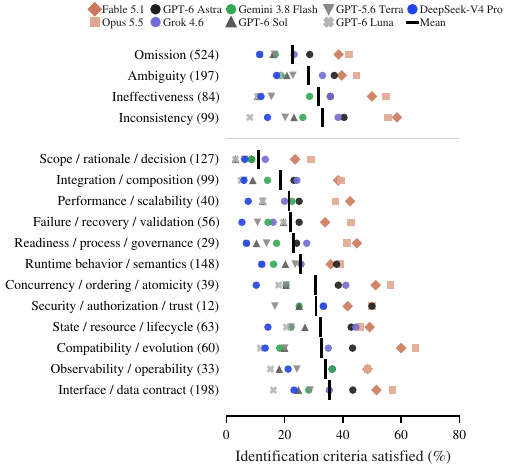}
    \caption{Identification-criterion satisfaction by deficiency class (top) and system concern area (bottom). Points show one labeled effort setting for each of nine models; bars show means over all 23 agents. Parentheses show category sizes.}
    \label{fig:category-performance}
\end{figure}

\textbf{Agents miss omissions most often.} We classify all specification deficiencies in our rubrics along two axes (Figure~\ref{fig:category-performance}): \emph{deficiency classes} (omission, ambiguity, inconsistency, and ineffectiveness) and \emph{system concern areas}. Among all deficiencies extracted from the historical RFCs, omissions make up the majority at 58.0\% (Appendix Figure~\ref{fig:rubric-taxonomy}). Yet across the four deficiency classes, agents perform worst on omissions, averaging only 22.8\% satisfaction (Figure~\ref{fig:category-performance}, top). Agents are more effective at checking statements written in a proposal for ambiguities or inconsistencies than at recognizing what the proposal fails to specify.

\textbf{Agents fail most on scope, rationale, and design decisions.} Across the twelve system concern areas (Figure~\ref{fig:category-performance}, bottom), agents achieve highest scores on localized engineering checks, such as backward compatibility and evolution (65.0\% for Opus 5.5 max) and interface and data contracts. By contrast, \emph{scope, rationale, and decisions} has the lowest satisfaction rate of any concern area, averaging 11.1\% across all agents, providing empirical evidence for practitioners' observations that current agents struggle with high-level design judgment~\citep{toub2026copilot,anthropic2025work}.

\begin{figure}[t]
    \centering
    \includegraphics[width=\columnwidth]{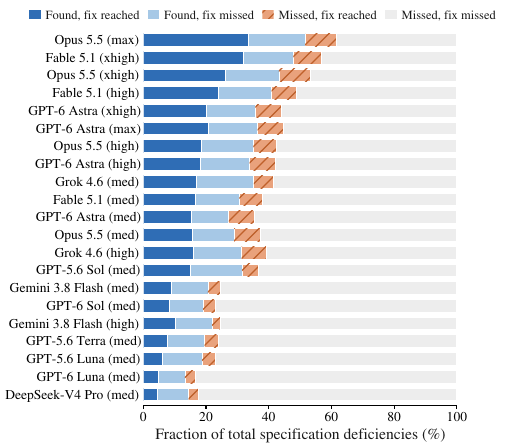}
    \caption{Breakdown of specification deficiencies correctly found vs. addressed by agents. \emph{Found, fix reached} indicates deficiencies that are both correctly identified and addressed.}
    \label{fig:find-fix}
\end{figure}

\textbf{Proposing revisions is harder than finding deficiencies.} Across all agents, revision scores are lower than identification scores (0.393 versus 0.489 for Opus 5.5 max; Appendix Table~\ref{tab:eval-models}). Figure~\ref{fig:find-fix} breaks down all 587 paired deficiencies in the rubrics by whether each agent finds the deficiency and whether it satisfies the corresponding revision criteria. Pooling across agents, only 47\% of revision criteria associated with correctly identified deficiencies are satisfied (and Opus 5.5 max satisfies only 60\%). Identifying a specification deficiency is therefore not sufficient: an agent must also be able to design revisions matching human expert level.

\begin{figure*}[t]
    \centering
    \includegraphics[width=\textwidth]{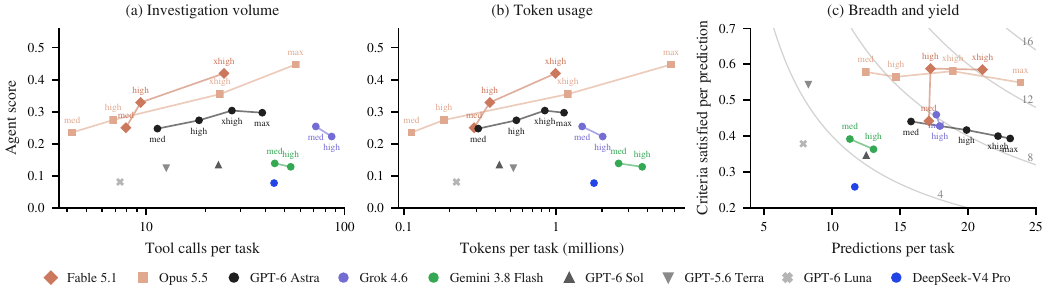}
    \caption{Score versus (a) tool calls and (b) tokens per task, and (c) predictions per task versus criteria satisfied per prediction, for nine models. Lines join one model's labeled reasoning-effort settings; gray curves in (c) mark constant criteria satisfied per task.}
    \label{fig:investigation}
\end{figure*}

\subsection{Agent Behavior and Dynamics}
\label{sec:what-separates}

Next, we study how agents interact with the benchmark environment and how their behaviors relate to their scores.

\textbf{More tool calls or tokens do not always lead to higher scores.} Across all agents, neither the number of tool calls nor the number of tokens consumed correlates with higher scores (Spearman $\rho=0.05$ for both; Figure~\ref{fig:investigation}a--b). The relationship varies by model family as reasoning effort increases: tool calls and tokens correlate positively with score for Claude models, show a weaker positive trend for OpenAI models, and correlate negatively for Gemini 3.8 Flash and Grok 4.6.

\begin{figure}[t]
    \centering
    \includegraphics[width=\columnwidth]{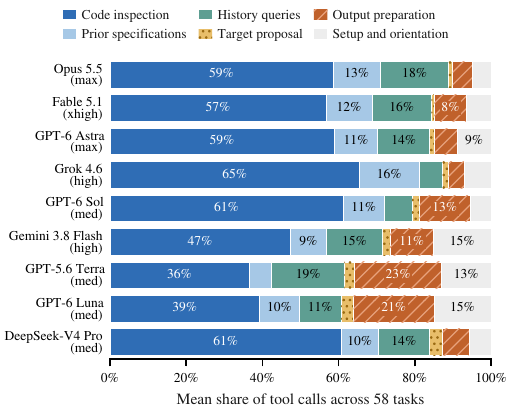}
    \caption{Mean share of tool calls by activity across all 58 tasks, weighting each task equally.}
    \label{fig:eval-trajectories}
\end{figure}

\textbf{Agents spend more tool calls inspecting code than consulting past RFCs.} We use an LLM to classify each tool call across all 58 tasks into activity categories (Figure~\ref{fig:eval-trajectories}). Code inspection is the largest activity for every agent (36--65\% of tool calls on average), whereas consulting prior RFCs and design discussions accounts for only 6--16\%. Because high-level design rationale, project conventions, and rejected alternatives are often recorded in past RFC discussions rather than in source code, we hypothesize that future agents could improve at specification-level reasoning by dedicating more exploration to prior specifications and design history.

\begin{figure}[t]
    \centering
    \includegraphics[width=\columnwidth]{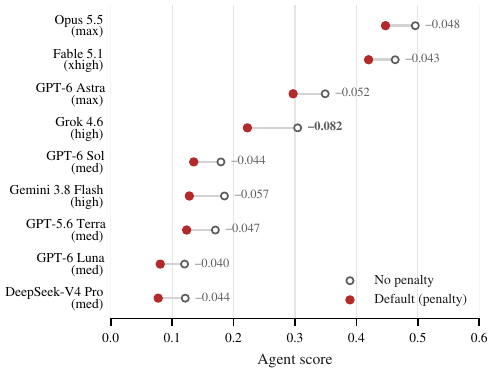}
    \caption{Agent score with the contradiction penalty (filled) and without (open). Labels show score lost to the penalty.}
    \label{fig:contradictions}
\end{figure}

\textbf{Submitting more predictions does not guarantee a higher score and increases contradictions.} Figure~\ref{fig:investigation}c plots the number of predictions an agent submits per task (\emph{breadth}) against the criteria satisfied per prediction (\emph{yield}). For example, Fable 5.1 (xhigh) submits fewer predictions per task than GPT-6 Astra (xhigh) (21.1 versus 22.2), but satisfies more criteria per prediction (0.59 versus 0.40) and achieves a higher score (0.420 versus 0.304). Furthermore, submitting more predictions risks contradicting decisions the experts agreed on (Figure~\ref{fig:contradictions}). Grok 4.6 (high) contradicts the most (2.1 per task, versus 1.2 for Opus 5.5 max), so the penalty lowers its score the most (from 0.304 to 0.223).

\subsection{Scoring Reliability}
\label{sec:judge-reliability}

Finally, we evaluate the reliability of the LLM judges.

\textbf{All nine LLM judgments agree on 85.79\% of binary decisions.} We measure agreement across the nine individual judgments for each of the 39,698 response--criterion pairs. Binary verdicts distinguish satisfied from not satisfied; ternary verdicts additionally distinguish contradicted from unsatisfied. Mean pairwise agreement is 94.08\% for binary verdicts and 92.28\% for ternary verdicts (Table~\ref{tab:eval-judges}). Fleiss' $\kappa$~\citep{fleiss1971agreement} is 0.836 and 0.819, respectively.

\begin{table}[t]
    \centering\footnotesize
    \setlength{\tabcolsep}{4pt}
    \caption{Agreement across all nine judging trials on 39,698 rubric criteria. Ternary labels account for contradictions.}
    \label{tab:eval-judges}

    \begin{tabular}{lrr}
\toprule
Metric & Binary & Ternary \\
\midrule
Mean pairwise agreement (\%) & 94.08 & 92.28 \\
Fleiss' $\kappa$ & 0.836 & 0.819 \\
Nine-way unanimity (\%) & 85.79 & 81.51 \\
\bottomrule
\end{tabular}

\end{table}

\textbf{LLM judges alignment with human reviewers.} To understand how LLM judges compare against human judges, two human reviewers independently labeled 28 anonymized model responses across 14 tasks (852 criterion evaluations; Appendix~\ref{app:human-agreement}). An individual LLM judge agrees with a human reviewer at the same rate that the two humans agree with each other: the two humans agree on 89.6\% of criteria (Cohen's $\kappa = 0.75$~\citep{cohen1960kappa}), while a single LLM judgment agrees with a human reviewer on 89.9\% on average across all 18 judge--human pairs ($\kappa = 0.74$; 91.3\% and 88.5\% for the two reviewers). Aggregating all nine LLM judgments into the panel majority raises agreement with the two human reviewers to 92.4\% and 89.7\% ($\kappa = 0.80$ and $0.74$; Appendix Table~\ref{tab:human-agreement}), and matches 95.8\% ($\kappa = 0.89$) of the 763 criteria where both humans agree.

\section{Related Work}
\label{sec:related-work}

\noindent\textbf{SWE Agent Benchmarks.}
The standard approach in prior SWE agent benchmarks is to provide the design specification as part of the task input and treat it as given. SWE-bench~\citep{jimenez2024swebench} established this paradigm by giving an agent a repository issue as its specification and grading the resulting code patch against tests, and subsequent benchmarks such as DeepSWE~\citep{huang2026deepswe} and SWE-Marathon~\citep{desai2026swemarathon} scale this setup to longer-horizon tasks. Terminal-Bench~\citep{merrill2026terminalbench} extends evaluation beyond repository patching to interactive command-line tasks in containerized environments, grading the final system state with test scripts. FrontierCode~\citep{cognition2026frontiercode} raises the bar from functional correctness to pull-request mergeability, using tests and maintainer-authored rubrics to grade test quality, scope discipline, and adherence to codebase conventions. ProgramBench~\citep{yang2026programbench} expands the scope from a patch to reconstructing an entire program from scratch given user-level documentation and a reference executable. While the agent must make internal design choices not detailed in the documentation during this process, ProgramBench's grading focuses on whether the rebuilt program matches the reference executable's behavior on hidden tests. Beyond code implementation, SWE Atlas~\citep{raghavendra2026sweatlas} evaluates a wider range of software engineering activities, including codebase question answering, test writing, and refactoring. Across all of the above benchmarks, the target behavior or existing system is treated as fixed and given, whereas \projectname{} evaluates whether a proposed specification is sound in the first place, before implementation.

\noindent\textbf{Requirement Clarification.}
A second line of work recognizes that user instructions in software engineering are often ambiguous or incomplete, and evaluates whether agents can clarify or recover missing requirements before writing code. Ambig-SWE~\citep{vijayvargiya2026ambigswe} and ClarifyCodeBench~\citep{fang2026clarifycodebench} withhold details from task prompts to test whether agents ask targeted clarifying questions in interactive dialogue with a simulated user before generating code, while SWE-RPG~\citep{zhou2026swerpg} evaluates whether agents recover implicit requirements from the repository as an intermediate step toward resolving GitHub issues. Whereas these benchmarks test whether agents can gather missing details to clarify an ambiguous coding request, \projectname{} tests whether agents can identify deficiencies and improve the specification itself.

\noindent\textbf{Proposal Judgment and Requirements Assessment.}
Finally, prior work on evaluating technical proposals and requirements either reduces the task to selecting among pre-written proposals or checks standalone requirements without a real codebase or project history. In SWE-Lancer's Manager tasks~\citep{miserendino2025swelancer}, an agent reviews competing implementation proposals submitted by freelancers for a job posting and picks the best one, which is graded against the historical manager's choice rather than requiring the agent to identify design flaws and propose revisions itself. Separately, requirements-engineering studies~\citep{lubos2024requirements,shefa2026requirements} prompt LLMs with standalone requirements and a brief project overview to check whether they satisfy standard quality guidelines (such as ISO 29148 or INCOSE), without a corresponding codebase. In contrast, \projectname{} evaluates open-ended specification review in real software systems, grounded in expert discussions that shaped consequential changes to those systems.

\section{Conclusion}
We present \projectname{}, a benchmark for evaluating \emph{specification-level reasoning} in software engineering agents. \projectname{} turns real Request for Comments (RFC) design proposals and expert discussions across seven production open-source systems into 58 tasks consisting of 1,726 human-verified rubric criteria. We evaluate 23 frontier SWE agents and show that the best-performing agent achieves a score of only 0.447 out of 1.0, and 35.6\% of expert-identified criteria are missed by every agent. As agents take on broader responsibility across the software engineering lifecycle, we argue that advancing the next generation of SWE agents requires evolving our benchmarks, models, and agent architectures beyond code implementation to specification-level reasoning.

\bibliography{references}
\bibliographystyle{mlsys2025}

\section*{AI Usage Disclosure}

\textbf{AI as part of the method.} LLMs are components of \projectname{}'s methodology, and the paper reports each use as part of the experimental design. LLMs draft, merge, refine, group, and weight rubric criteria before and after human review (Section~\ref{sec:impl-rubrics}); an ensemble of three LLM judges scores agent predictions (Section~\ref{sec:impl-judging}); and GPT-6.1 Sol labels criteria by deficiency class and concern area (Section~\ref{sec:where-fail}) and classifies agent tool calls (Section~\ref{sec:what-separates}). The 23 evaluated agent configurations are the subject of the experiments.

\textbf{AI assistance in producing the work.} We used LLM-based coding agents (Claude Code, Codex, and Antigravity) in four parts of the work. \emph{Ideation and experiment design:} the agents proposed and critiqued alternative benchmark and experiment designs; the authors made all design decisions. \emph{Literature search:} the agents located and summarized candidate related work and benchmark documentation; the authors read every cited source and verified every citation. \emph{Code:} the agents wrote parts of the benchmark harness, analysis scripts, and plotting code; the authors reviewed and tested this code. \emph{Writing:} the agents drafted portions of the text from the authors' outlines and notes and proposed revisions; the authors rewrote and checked all text and verified every number, claim, and reference against the benchmark data and sources. The authors take full responsibility for the content of this paper.

\clearpage
\appendix
\let\addcontentsline\origaddcontentsline
\setcounter{tocdepth}{2}
\etocsettocstyle{\section*{Appendix Contents}}{}
\tableofcontents

\section{Coding Benchmarks in Frontier Model Releases}
\label{app:benchmark-survey}

We surveyed model releases from Anthropic, DeepSeek, Google, Meta, OpenAI, and xAI between October 21, 2025 and October 21, 2026. We included flagship and mid-tier general-purpose models and excluded small models, snapshot updates, restricted previews, and specialized models. From each release's announcement, model card, and system card, we recorded every public benchmark reported among its coding results, merging versions of the same benchmark. We excluded internal benchmarks, results quoted from partners, safety evaluations, and composite indices. Thirty-nine releases reported at least one public coding benchmark: 12 from OpenAI, 11 from Anthropic, 6 from Meta, 4 from DeepSeek, 3 from Google, and 3 from xAI.

A benchmark \emph{fixes the required behavior} if each task states the behavior in advance, as an issue, task description, reference program, or problem statement, and grading checks whether the agent's output achieves it. Table~\ref{tab:benchmark-survey} classifies the 26 benchmarks from their primary sources. Twenty-four fix the required behavior. SWE-Atlas Codebase QnA grades answers to questions about a codebase, and CursorBench's private tasks also include code review and codebase understanding. No benchmark's public documentation describes tasks that ask whether the required behavior is itself sound.

\begin{table*}[t]
\centering\scriptsize
\setlength{\tabcolsep}{4pt}
\caption{Public coding benchmarks reported in 39 frontier model releases from October 2025 to October 2026. \emph{Releases} counts the releases that report each benchmark, with the number of developers in parentheses.}
\label{tab:benchmark-survey}
\begin{tabularx}{\textwidth}{@{}>{\raggedright\arraybackslash}p{120pt}r>{\raggedright\arraybackslash}X>{\raggedright\arraybackslash}p{120pt}>{\raggedright\arraybackslash}p{60pt}@{}}
\toprule
Benchmark & Releases & Agent receives & Graded by & Fixes behavior \\
\midrule
Terminal-Bench~\citep{merrill2026terminalbench} & 34 (6) & Task instruction and a container & Tests on the final container state & Yes \\
SWE-bench Pro~\citep{deng2025swebenchpro} & 20 (6) & Repository, issue, requirements, and interface & Fail-to-pass and pass-to-pass tests & Yes \\
DeepSWE~\citep{huang2026deepswe} & 18 (6) & Repository and a description of functionality to build & Hand-written behavioral verifier & Yes \\
SWE-bench Verified~\citep{chowdhury2024swebenchverified} & 16 (5) & Repository and issue & Tests from the resolving pull request & Yes \\
SWE-bench Multilingual~\citep{yang2025swesmith} & 14 (3) & Repository and issue, nine languages & Tests from the resolving pull request & Yes \\
FrontierCode~\citep{cognition2026frontiercode} & 9 (3) & Repository and a change to make & Tests and rubrics for mergeability & Yes \\
ProgramBench~\citep{yang2026programbench} & 9 (3) & Executable and its documentation & Behavioral tests against the executable & Yes \\
SWE-bench Multimodal~\citep{yang2024swebenchmultimodal} & 8 (1) & Repository and issue with images & Unit tests & Yes \\
CursorBench~\citep{cursor2026cursorbench} & 8 (2) & Private tasks from real Cursor sessions & Agentic graders & Mixed: also code review and comprehension \\
FrontierSWE~\citep{proximal2026frontierswe} & 7 (3) & Long-horizon engineering or research task & Hidden verifiers and reference baselines & Yes \\
SWE-Marathon~\citep{desai2026swemarathon} & 4 (2) & Outcome-level instructions and an environment & Hidden multi-layer verifiers & Yes \\
SWE-Lancer (IC SWE)~\citep{miserendino2025swelancer} & 3 (1) & Repository and freelance task & End-to-end tests & Yes \\
LiveCodeBench Pro~\citep{zheng2025livecodebenchpro} & 3 (2) & Contest problem & Online judge with hidden tests & Yes \\
LiveCodeBench~\citep{jain2024livecodebench} & 3 (1) & Contest problem & Test cases & Yes \\
Codeforces~\citep{openai2025competitive} & 3 (1) & Contest problem & Platform judge with hidden tests & Yes \\
Vibe Code Bench~\citep{tran2026vibecodebench} & 2 (2) & Web application specification & Browser agent runs user workflows & Yes \\
APEX-SWE~\citep{kottamasu2026apexswe} & 2 (1) & Integration or observability task & Hidden tests & Yes \\
SWE-Atlas Codebase QnA~\citep{raghavendra2026sweatlas} & 2 (1) & Repository and a question about it & LLM judge with expert rubrics & No: code comprehension \\
Aider Polyglot~\citep{aider2024polyglot} & 1 (1) & Exercise instructions and stub files & Unit tests & Yes \\
NL2Repo-Bench~\citep{ding2025nl2repo} & 1 (1) & Requirements document & The original project's tests & Yes \\
SciCode~\citep{tian2024scicode} & 1 (1) & Function header and docstring & Scientist-written tests & Yes \\
WebDev Arena~\citep{vichare2025webdevarena} & 1 (1) & User prompt for a web application & Human preference votes & Yes \\
HumanEval~\citep{chen2021humaneval} & 1 (1) & Function signature and docstring & Unit tests & Yes \\
BigCodeBench~\citep{zhuo2025bigcodebench} & 1 (1) & Function signature and docstring & Unit tests & Yes \\
IOI 2025~\citep{ioi2025} & 1 (1) & Olympiad tasks & Official grader & Yes \\
ICPC World Finals 2025~\citep{icpc2025} & 1 (1) & Contest problems & Official judge & Yes \\
\bottomrule
\end{tabularx}
\end{table*}

\section{Tasks and Agent Environments}
\label{app:tasks-environments}

\subsection{Task List}
\label{app:tasks}

Table~\ref{tab:tasks} lists the 58 tasks, which span proposals from 2014 to 2025. Each identifier is the proposal's pull request in the project's RFC repository for Kubernetes, React, Rust, and TVM, its RFC issue for vLLM, its PEP number for Python, and its patch series for Linux. The date is the task cutoff, when the initial proposal was submitted. The 58 rubrics contain 1,726 criteria, 904 deficiency and 822 revision criteria, with 6 to 71 criteria per task (median 28.5).

\begin{table*}[t]
\centering\scriptsize
\setlength{\tabcolsep}{3pt}\renewcommand{\arraystretch}{0.92}
\caption{All 58 tasks. \emph{ID} identifies the proposal's review record; \emph{Date} is the task cutoff; \emph{D} and \emph{R} count deficiency and revision criteria.}
\label{tab:tasks}
\begin{tabular}[t]{@{}rlrrr@{}}
\toprule
ID & Proposal & Date & D & R \\
\midrule
\multicolumn{5}{@{}l}{\textbf{Kubernetes} (enhancements PR)} \\
4861 & KEP-4817: Resource claim device status & 2024-09 & 32 & 39 \\
5287 & KEP-5278: Nominated node name for expected placement & 2025-05 & 19 & 23 \\
5331 & KEP-5325: HPA pod selection accuracy & 2025-05 & 17 & 19 \\
5382 & KEP-5381: Mutable PersistentVolume node affinity & 2025-06 & 17 & 20 \\
5510 & KEP-5501: PreEnqueue rejections in pod status & 2025-09 & 16 & 17 \\
5514 & KEP-5468: Invariant testing & 2025-09 & 9 & 9 \\
5556 & KEP-5532: RestartAllContainers action & 2025-09 & 22 & 29 \\
5558 & KEP-4671: Gang scheduling & 2025-09 & 36 & 34 \\
\midrule
\multicolumn{5}{@{}l}{\textbf{vLLM} (RFC issue)} \\
2614 & Automatic prefix caching & 2024-01 & 8 & 2 \\
3587 & Distributed inference environment interface & 2024-03 & 13 & 15 \\
5723 & Runtime weight update API & 2024-06 & 11 & 6 \\
6077 & Priority scheduling & 2024-07 & 10 & 9 \\
8967 & Reward models in the OpenAI-compatible server & 2024-09 & 9 & 5 \\
11945 & Pipeline parallelism for vLLM V1 & 2025-01 & 12 & 5 \\
12249 & Hidden states processor & 2025-01 & 9 & 3 \\
14764 & Input shape schema for multimodal models & 2025-03 & 8 & 9 \\
\midrule
\multicolumn{5}{@{}l}{\textbf{React} (rfcs PR)} \\
6 & Async-safe static lifecycle hooks & 2017-12 & 12 & 13 \\
30 & Ref forwarding API & 2018-03 & 15 & 14 \\
51 & Profiler & 2018-05 & 21 & 16 \\
63 & \texttt{React.memo} & 2018-10 & 13 & 7 \\
64 & \texttt{React.lazy} & 2018-10 & 9 & 7 \\
147 & \texttt{useMutableSource} & 2020-02 & 22 & 23 \\
188 & Server Components & 2020-12 & 35 & 22 \\
213 & Suspense in React 18 & 2022-03 & 10 & 10 \\
\midrule
\multicolumn{5}{@{}l}{\textbf{Python} (PEP)} \\
584 & Union operators for \texttt{dict} & 2019-03 & 12 & 18 \\
616 & String methods to remove prefixes and suffixes & 2020-03 & 6 & 5 \\
618 & Optional length checking for \texttt{zip} & 2020-05 & 3 & 3 \\
695 & Type parameter syntax & 2022-07 & 14 & 12 \\
696 & Type defaults for type variables & 2022-07 & 22 & 14 \\
701 & Syntactic formalization of f-strings & 2022-11 & 8 & 9 \\
702 & Type system support for deprecations & 2022-12 & 12 & 12 \\
705 & \texttt{TypedMapping} & 2023-02 & 18 & 12 \\
\bottomrule
\end{tabular}\hfill
\begin{tabular}[t]{@{}rlrrr@{}}
\toprule
ID & Proposal & Date & D & R \\
\midrule
\multicolumn{5}{@{}l}{\textbf{Rust} (rfcs PR)} \\
195 & Associated items and multidispatch & 2014-08 & 28 & 14 \\
2388 & Reserve \texttt{try} for \texttt{try} blocks & 2018-04 & 8 & 10 \\
2788 & Standard lazy types & 2019-10 & 17 & 13 \\
3519 & Arbitrary self types v2 & 2023-11 & 13 & 13 \\
3537 & MSRV-aware Cargo dependency resolution & 2023-11 & 25 & 21 \\
3668 & Async closures & 2024-07 & 19 & 18 \\
3791 & \texttt{--crate-attr} compiler flag & 2025-03 & 10 & 8 \\
3809 & \texttt{derive(From)} for single-field structs & 2025-05 & 7 & 6 \\
\midrule
\multicolumn{5}{@{}l}{\textbf{TVM} (tvm-rfcs PR)} \\
9 & Unified static memory planning & 2021-07 & 22 & 22 \\
22 & TIR non-scalar constants & 2021-08 & 24 & 14 \\
23 & Annotation field for \texttt{tir.allocate} & 2021-08 & 8 & 9 \\
25 & PyTorchTVM & 2021-08 & 13 & 8 \\
39 & Layout transformations on buffer access & 2021-10 & 28 & 27 \\
60 & Universal Modular Accelerator Interface & 2022-03 & 25 & 24 \\
62 & Collage & 2022-03 & 23 & 17 \\
77 & Buffer layout padding & 2022-06 & 23 & 25 \\
79 & TVMScript metaprogramming & 2022-06 & 25 & 30 \\
100 & SparseTIR dialect & 2023-04 & 16 & 13 \\
\midrule
\multicolumn{5}{@{}l}{\textbf{Linux} (patch series)} \\
mm & Large-folio swap-in refault handling & 2024-04 & 16 & 14 \\
slub & Delay freezing of CPU partial slabs & 2023-10 & 16 & 17 \\
tsm & Runtime measurement registers ABI & 2024-01 & 12 & 8 \\
erofs & Shared domain & 2022-08 & 17 & 16 \\
vsock & Credit update logic & 2023-03 & 6 & 8 \\
PCI & PCIe device tree support for TI J784S4 & 2024-01 & 5 & 7 \\
perf & Arm SPE process and thread tracking & 2021-11 & 10 & 12 \\
regulator & Regulator output connector binding & 2022-09 & 8 & 7 \\
\midrule
\multicolumn{5}{@{}l}{\textbf{Total} (58 tasks)} \\
 & & & 904 & 822 \\
\bottomrule
\end{tabular}
\end{table*}

\subsection{Environment Contents}
\label{app:environments}

Each environment contains the project's code repository and, where one exists, its RFC repository, checked out at a commit from before the cutoff with the Git history up to that commit (Table~\ref{tab:env-coverage}). For projects that discuss RFCs on GitHub, it also contains a saved copy of past RFC issues and pull requests, which the agent searches offline with a command-line tool. Linux environments contain no discussion records because we do not include mailing-list archives.

\begin{table}[t]
\centering\scriptsize
\setlength{\tabcolsep}{3pt}
\caption{Repositories and offline discussion records in each project's environment. All repositories include Git history up to the cutoff.}
\label{tab:env-coverage}
\begin{tabular}{@{}lll@{}}
\toprule
Project & Repositories & Offline discussions \\
\midrule
Kubernetes & \texttt{kubernetes}, \texttt{enhancements} & KEP issues and PRs \\
React & \texttt{react}, \texttt{rfcs} & RFC issues and PRs \\
Rust & \texttt{rust}, \texttt{rfcs} & RFC issues and PRs \\
TVM & \texttt{tvm}, \texttt{tvm-rfcs} & RFC PRs \\
Python & \texttt{cpython}, \texttt{peps} & PEP issues and PRs \\
vLLM & \texttt{vllm} & RFC-labeled issues \\
Linux & \texttt{linux} & None \\
\bottomrule
\end{tabular}
\end{table}

\textbf{Removing later information.} The environment builder deletes every Git object that is not reachable from the selected commit, along with all other branches, tags, remotes, and reflogs, and verifies that every remaining commit predates the cutoff. From the saved discussions, it removes the target proposal's record and every record created or updated after the cutoff. Each run starts from a fresh copy of the environment, and the runner repeats these checks before launching the agent.

\subsection{Agent Prompt}
\label{app:agent-prompt}

Each agent receives one prompt per task: the task instructions, the workspace rules, a description of the project's environment, and the initial proposal. Figure~\ref{fig:agent-prompt} reproduces the prompt for Kubernetes tasks. Prompts for other projects differ only in the repository names and the available discussion records.

\begin{figure*}[p]
\begin{tcolorbox}[promptbox,unbreakable,title={Agent prompt for Kubernetes tasks}]
You are an expert maintainer of this project. You are given access to the project codebase and an initial Request for Comments (RFC) proposal.

Your task is to identify deficiencies in the initial RFC proposal. This includes but is not limited to: ambiguities, incompleteness, inconsistencies, etc.

For each deficiency you identify, also provide the most likely concrete revision that would address it. The revision text must naturally explain why that change is appropriate for resolving the deficiency. Separately provide supporting evidence from the local repository/RFC/history.

\textbf{\# Prediction Rules}
\begin{itemize}[nosep,leftmargin=*]
\item Produce every distinct, substantive predicted pair you can support. There is no fixed prediction-count budget.
\item Rank pairs by importance and severity. Put the most critical pairs first.
\item The \texttt{revision} field must be a concrete RFC/spec change, not merely a request for more discussion, documentation, or implementation work.
\item Include the reason the change is appropriate naturally within the \texttt{revision} text. Do not add a separate reason column or require a literal label. Do not use the \texttt{evidence} field as a substitute for explaining the design rationale.
\item Use the codebase, project history, prior RFCs, and the initial RFC to infer likely reviewer decisions. Prefer the smallest concrete spec/RFC change supported by evidence.
\end{itemize}

\textbf{\# Access Rules}
\begin{itemize}[nosep,leftmargin=*]
\item Do not use or assume access to the target RFC's later GitHub review thread. Do not search for or read that target discussion even if repository history tooling could locate it; that discussion contains the benchmark answer key.
\item Do not inspect or cite commits, branches, tags, pull requests, issues, files, or generated artifacts created after the task cutoff. If you use git, restrict yourself to the current checked-out HEAD and history reachable from HEAD. Do not use \texttt{git log --all}, remote refs, tags, reflogs, dangling commits, or arbitrary commit hashes to find future implementation/resolution commits.
\end{itemize}

\textbf{\# Output}

Write exactly one CSV file under the starting working directory: \texttt{predicted\_pairs.csv} with exactly these columns:
\begin{itemize}[nosep,leftmargin=*]
\item id: integer index starting from 1
\item subject: affected spec/system subject
\item deficiency: deficiency in the initial RFC
\item revision: concrete RFC/spec change that resolves this deficiency, including a natural explanation of why that change is appropriate
\item evidence: concise supporting evidence from the codebase, RFC, or historical context
\end{itemize}
[Instructions on derived files, RFC~4180 CSV quoting, and a two-row example CSV omitted.]

Your current working directory is \texttt{\{cwd\}}. You cannot write to directories outside this directory. You do not have access to the internet.

\textbf{\# IMPORTANT: Workspace Boundary}

Stay strictly within the current task workspace, \texttt{\{cwd\}}, and its subdirectories.
\begin{itemize}[nosep,leftmargin=*]
\item Do not access benchmark infrastructure, task definitions, evaluation data, gold answers, other runs, or other agents' workspaces.
\item Do not use the network.
\end{itemize}
[Rules against leaving the workspace through parent or absolute paths, for running Git only on the documented repositories, and for reporting missing resources omitted.]

\textbf{\# Directory Structure and Past RFC History}

The \texttt{kubernetes/} repo contains the source code of Kubernetes. You can use \texttt{git} cli tool to access its git history. Github history (PRs/issues) are not available.

The \texttt{rfcs/} repo contains historical Kubernetes RFCs. Use this repo to understand how past RFCs were discussed and accepted.

You can use \texttt{git} cli tool to access its git history. You can use the \texttt{get\_gh\_history.py} script in that directory to access its Github history (PRs and issues):
\begin{Verbatim}[fontsize=\scriptsize]
python get_gh_history.py issue --id ID
python get_gh_history.py pr --keyword KEYWORD
\end{Verbatim}
\noindent[Equivalent commands for the other record type omitted.]

\textbf{\# The RFC content:}\\
{}[Full text of the initial proposal.]
\end{tcolorbox}
\caption{The agent prompt for Kubernetes tasks. \texttt{\{cwd\}} is the task workspace; brackets abbreviate four mechanical passages.}
\label{fig:agent-prompt}
\end{figure*}

\section{Rubric Construction}
\label{app:rubric-construction}

Table~\ref{tab:rubric-pipeline} lists the nine stages that implement the four steps of Section~\ref{sec:impl-rubrics}. Models draft, refine, and organize the criteria; researchers make the final decision on every criterion and every weight. No stage sees agent predictions.

\begin{table}[t]
\centering\scriptsize
\setlength{\tabcolsep}{3pt}
\caption{Rubric construction stages.}
\label{tab:rubric-pipeline}
\begin{tabular}{@{}rp{3.1cm}p{3.6cm}@{}}
\toprule
& Stage & Performed by \\
\midrule
1 & Draft criteria independently & GPT-5.6 Sol, Claude Sonnet 5, Gemini 3.6 Flash \\
2 & Merge drafts by meaning & GPT-5.6 Sol \\
3 & Review criterion quality & GPT-5.6 Sol \\
4 & Revise from quality review & GPT-5.6 Sol \\
5 & Review every criterion & Two researchers \\
6 & Group criteria into a tree & GPT-5.6 Sol \\
7 & Weight nodes independently & GPT-5.6 Sol, Claude Opus 5, Gemini 3.7 Flash \\
8 & Merge weights by majority vote & Deterministic code \\
9 & Inspect weights, break ties & Researchers \\
\bottomrule
\end{tabular}
\end{table}

\subsection{Extracting and Refining Criteria}
\label{app:rubric-extraction}

Each drafting model receives the initial proposal and the expert discussion and writes a list of yes/no criteria that meet five requirements:
\begin{itemize}[nosep,leftmargin=*]
    \item \textbf{Two types.} A deficiency criterion tests whether the agent identifies a deficiency the experts raised; a revision criterion tests whether it proposes a resolution the experts established.
    \item \textbf{Grounding.} Every criterion traces to the discussion. A suggestion or open alternative becomes a revision criterion only if the discussion settles on it.
    \item \textbf{Resolution status.} Revision criteria cover only what the discussion resolved, so an unresolved deficiency receives no revision criterion.
    \item \textbf{Atomicity.} Each criterion tests one deficiency or one decision that an agent could recover independently of the others.
    \item \textbf{Judgeability.} Each criterion is self-contained and states which near misses earn no credit.
\end{itemize}

A model merges the three drafts by meaning. A second pass scores each merged criterion from 1 to 5 on ten quality dimensions (Table~\ref{tab:quality-dimensions}) and recommends changes. A third pass applies a recommendation only if the proposal and discussion support it.

\begin{table}[t]
\centering\scriptsize
\setlength{\tabcolsep}{3pt}
\caption{Quality dimensions for reviewing criteria (stage 3).}
\label{tab:quality-dimensions}
\begin{tabular}{@{}p{2.0cm}p{5.6cm}@{}}
\toprule
Dimension & A criterion scores 5 if it \ldots \\
\midrule
Source grounding & is explicitly established by the proposal and the expert discussion \\
Stage fidelity & tests exactly its declared stage, deficiency or revision \\
Resolution status & matches how far the discussion resolved the issue \\
Issue fidelity & preserves the experts' deficiency or decision, its scope, and its rationale \\
Substantive value & matters to behavior, compatibility, safety, integration, or acceptance \\
Atomicity & captures one deficiency or one independently recoverable decision \\
Specificity & states its subject, condition, and boundary without outside context \\
Instruction form & tells the judge what the agent must identify or propose \\
Binary judgeability & lets a judge reliably decide yes or no, including near misses \\
Distinct contribution & can be satisfied independently of every other criterion \\
\bottomrule
\end{tabular}
\end{table}

\subsection{Human Review}
\label{app:human-review}

Two graduate-level researchers each review all 1,709 criteria from stage 4 independently. For each criterion, a reviewer reads the proposal and the relevant discussion and chooses to keep, revise, split, merge, or drop it. The reviewers have no written guidelines beyond the criterion requirements above, and they resolve every disagreement by discussion. In their independent passes, both keep the same 1,314 criteria (76.9\%) unchanged. For example, they dropped a vLLM revision criterion that required an output mechanism the discussion had only suggested. Section~\ref{sec:impl-rubrics} reports the final outcomes.

\subsection{Tree Construction and Weighting}
\label{app:tree-weighting}

A model groups each task's criteria into a tree of at most four levels without changing any criterion. Each criterion is exactly one leaf, and internal nodes only group criteria. Three models then independently weight every node 1, 2, or 3 by its importance relative to its siblings, given the tree, the proposal, and the discussion. A node's weight reflects the importance of its subtree, not its number of leaves. Of the 2,481 non-root nodes, all three models agree on 1,200 (48.4\%) and two agree on 1,249 (50.3\%). Researchers break the 32 three-way ties and inspect every majority weight.

\subsection{Rubric Taxonomy}
\label{app:rubric-taxonomy}

\begin{figure}[t]
    \centering
    \includegraphics[width=\columnwidth]{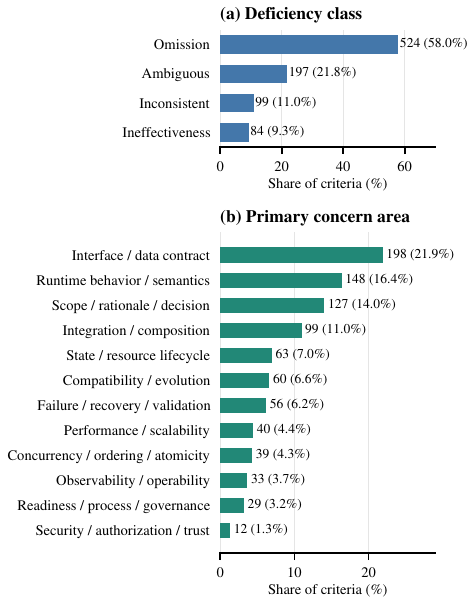}
    \caption{Composition of the 904 deficiency criteria by deficiency class and concern area.}
    \label{fig:rubric-taxonomy}
\end{figure}

To analyze where agents fail (Section~\ref{sec:where-fail}), GPT-6.1 Sol labels each criterion with a deficiency class and a concern area, given the proposal and the task's rubric. The four deficiency classes are:
\begin{itemize}[nosep,leftmargin=*]
    \item \textbf{Omission:} the proposal does not cover a necessary case, behavior, interface, constraint, or justification.
    \item \textbf{Ambiguity:} the proposal's text permits materially different reasonable interpretations.
    \item \textbf{Inconsistency:} the proposal conflicts with itself or with an established interface, invariant, or design principle of the system.
    \item \textbf{Ineffectiveness:} the specified behavior, even if implemented faithfully, fails to achieve the proposal's goal.
\end{itemize}
The twelve concern areas name the part of the system that the deficiency affects (Figure~\ref{fig:rubric-taxonomy}). Omissions make up 58.0\% of the 904 deficiency criteria, and interface and data contracts (21.9\%) is the largest concern area.

\section{Judging and Scoring}
\label{app:judging}

\subsection{Judge Configuration and Prompt}
\label{app:judge-prompt}

The three judges run at medium reasoning effort without tools. Each judge receives the prompt in Figure~\ref{fig:judge-prompt}, the task's rubric as JSON, and the agent's prediction file. For every criterion, it returns one of five verdicts, up to three supporting prediction identifiers, and a justification of at most 40 words.

\begin{figure*}[t]
\begin{tcolorbox}[promptbox,unbreakable,title={Judge prompt}]
You are the independent binary rubric judge for \projectname{}.

Evaluate every rubric item exactly once against the supplied pair-first predictions. The rubric item is the authoritative credit target.

\textbf{Rules:}
\begin{itemize}[nosep,leftmargin=*]
\item For gap\_identification, the prediction's subject/deficiency must identify the required original gap. A correct revision cannot compensate for a missing gap.
\item For revision, the prediction's revision must state the required concrete resolution for the same issue. A correct gap cannot compensate for a vague or incompatible revision.
\item Semantic equivalence is sufficient, but never infer content the prediction does not state. Evidence and plausible intent do not replace prediction text.
\item One prediction may support multiple items, and multiple predictions may collectively satisfy one item.
\item Scoring is binary. Partial or adjacent coverage scores 0.
\end{itemize}

\textbf{Verdicts:}
\begin{itemize}[nosep,leftmargin=*]
\item SATISFIED: score 1 and at least one SUPPORTS reference.
\item NO\_RELEVANT\_PREDICTION: score 0 and no references.
\item PARTIAL\_OR\_INCOMPLETE: score 0 and at least one PARTIAL reference.
\item CONTRADICTED: score -1 and at least one CONTRADICTS reference.
\item RELATED\_BUT\_NOT\_SATISFIED: score 0 and at least one RELATED reference.
\end{itemize}

Reference only prediction IDs supplied below. For each rubric item, return its verdict category, zero to three relevant prediction IDs, and one criterion-specific justification of at most 40 words. Do not repeat the same reason separately for every prediction. The runner derives the binary score, reference relationship, summary, and empty discussion locations.

Do not call tools, inspect files, browse, or use the network. All required inputs are contained in this message. Return only the schema-conforming JSON.

\textcolor{gray}{=== RUBRIC JSON ===}\\
{}[Reviewed rubric of the task.]\\
\textcolor{gray}{=== PREDICTED PAIRS CSV ===}\\
{}[Agent's prediction file.]
\end{tcolorbox}
\caption{The judge prompt. The bracketed lines stand for the task's rubric and the agent's predictions.}
\label{fig:judge-prompt}
\end{figure*}

\subsection{Score Definitions}
\label{app:score-definition}

\textbf{Criterion score.} We map each of the nine verdicts for criterion $\ell$ to $+1$ for SATISFIED, $-1$ for CONTRADICTED, and $0$ otherwise. The criterion score $s_\ell$ is the median of the nine values.

\textbf{Node and task scores.} Each non-root node $c$ has a weight $w_c \in \{1,2,3\}$. The score of an internal node $n$ is the weighted mean of its children's scores,
\[
s_n = \frac{\sum_{c \in \mathrm{children}(n)} w_c\, s_c}{\sum_{c \in \mathrm{children}(n)} w_c}.
\]
The task score is the root's score and lies in $[-1, 1]$. An agent's benchmark score is the mean of its 58 task scores.

\textbf{Identification and revision scores.} Expanding the recursion writes the task score as $\sum_\ell \pi_\ell s_\ell$. The share $\pi_\ell$ is a product over the nodes on the path from the root to $\ell$: for each node, its weight divided by the total weight of it and its siblings. The identification score restricts the sum to deficiency criteria $D$ and renormalizes, $\sum_{\ell \in D} \pi_\ell s_\ell / \sum_{\ell \in D} \pi_\ell$. The revision score does the same for revision criteria. Appendix~\ref{app:gang-score} works through one task.

\section{Additional Results}
\label{app:evaluation-details}

\subsection{Results for All Agents}

Table~\ref{tab:eval-models} reports the scores, tokens, and API cost of all 23 agents.

\begin{table}[t]
    \centering\scriptsize
    \setlength{\tabcolsep}{3pt}
    \caption{Results for all 23 agents. Appendix~\ref{app:score-definition} defines the identification and revision scores. Tokens and cost are means per task.}
    \label{tab:eval-models}
    \begin{tabular}{lrrrrr}
\toprule
Agent & Score & Identify & Revise & Tokens (M) & API USD \\
\midrule
Opus 5.5 (max) & 0.447 & 0.489 & 0.393 & 5.695 & 4.888 \\
Fable 5.1 (xhigh) & 0.420 & 0.468 & 0.370 & 0.992 & 3.465 \\
Opus 5.5 (xhigh) & 0.355 & 0.408 & 0.294 & 1.193 & 1.514 \\
Fable 5.1 (high) & 0.329 & 0.384 & 0.282 & 0.367 & 1.718 \\
GPT-6 Astra (xhigh) & 0.304 & 0.353 & 0.245 & 0.843 & 2.554 \\
GPT-6 Astra (max) & 0.297 & 0.353 & 0.224 & 1.127 & 3.759 \\
Opus 5.5 (high) & 0.274 & 0.330 & 0.214 & 0.183 & 0.470 \\
GPT-6 Astra (high) & 0.273 & 0.329 & 0.209 & 0.548 & 1.831 \\
Grok 4.6 (med) & 0.254 & 0.342 & 0.166 & 1.482 & 1.271 \\
Fable 5.1 (med) & 0.250 & 0.306 & 0.190 & 0.289 & 1.518 \\
GPT-6 Astra (med) & 0.247 & 0.280 & 0.200 & 0.308 & 1.043 \\
Fable 5 (med) & 0.238 & 0.290 & 0.182 & 0.319 & 1.684 \\
Opus 5.5 (med) & 0.235 & 0.275 & 0.191 & 0.112 & 0.317 \\
Grok 4.6 (high) & 0.223 & 0.309 & 0.140 & 2.023 & 1.517 \\
GPT-5.6 Sol (med) & 0.206 & 0.295 & 0.121 & 0.817 & 0.818 \\
Gemini 3.8 Flash (med) & 0.139 & 0.199 & 0.072 & 2.574 & 0.556 \\
GPT-6 Sol (med) & 0.135 & 0.198 & 0.068 & 0.424 & 0.229 \\
Gemini 3.8 Flash (high) & 0.128 & 0.213 & 0.035 & 3.679 & 0.895 \\
GPT-5.6 Terra (med) & 0.124 & 0.191 & 0.057 & 0.524 & 0.292 \\
GPT-5.6 Luna (med) & 0.106 & 0.193 & 0.021 & 0.435 & 0.027 \\
GPT-6 Luna (med) & 0.081 & 0.138 & 0.024 & 0.221 & 0.006 \\
DeepSeek-V4 Pro (med) & 0.077 & 0.146 & 0.008 & 1.774 & 0.449 \\
Gemini 3.1 Pro (high) & 0.043 & 0.069 & 0.015 & 1.046 & 0.756 \\
\bottomrule
\end{tabular}

\end{table}

\textbf{Accounting.} Total tokens include cached input, and output tokens include reasoning tokens. Costs apply each provider's standard short-context rates as of October 10, 2026 (peak rates for DeepSeek) to the recorded usage. They exclude long-context surcharges, cache storage, subscriptions, and judging.

\subsection{Scoring Sensitivity}
\label{app:scoring-sensitivity}

We rescore the saved verdicts under four alternatives to the default scoring and compare the resulting agent rankings with the default ranking. Removing the contradiction penalty raises agent scores by 0.026--0.082 (Figure~\ref{fig:contradictions}) with a Spearman rank correlation of $\rho=0.985$. Averaging the nine verdicts instead of taking their median gives $\rho=0.997$. Replacing the tree weights with uniform weights across criteria gives $\rho=0.993$, and giving siblings equal weights gives $\rho=0.998$.

\subsection{Agent Trajectory Annotation}
\label{app:trajectories}

For Figure~\ref{fig:eval-trajectories}, GPT-6.1 Sol labels every tool invocation in 522 runs (nine agents on all 58 tasks) with one activity. It sees each invocation's arguments, an excerpt of its result, and the neighboring steps. The six categories are:
\begin{itemize}[nosep,leftmargin=*]
    \item \textbf{Target proposal:} reading or searching the assigned proposal.
    \item \textbf{Prior specifications:} reading earlier RFCs, design documents, or design guidance.
    \item \textbf{Code inspection:} exploring source code, tests, configuration, or documentation, including running small probes.
    \item \textbf{History query:} inspecting Git history or past issue and pull request discussions.
    \item \textbf{Output preparation:} writing or checking the prediction file.
    \item \textbf{Other:} workspace orientation, setup, and steps without a clearer purpose.
\end{itemize}

\subsection{Agreement Across Nine Judgments}
\label{app:nine-judgment-agreement}

For instance $i$, let $n_{ic}$ count the judgments assigning label $c$, with $\sum_c n_{ic}=9$. Its pairwise agreement is $P_i=\sum_c n_{ic}(n_{ic}-1)/72$. For example, an 8--1 binary split gives $P_i=28/36$, whereas a 5--4 split gives $P_i=16/36$. We report $\bar P=N^{-1}\sum_i P_i$, where $N=39{,}698$. Let $p_c=(9N)^{-1}\sum_i n_{ic}$ be the pooled frequency of label $c$. Fleiss' $\kappa=(\bar P-P_e)/(1-P_e)$ uses $P_e=\sum_c p_c^2$ as its chance baseline. We treat labels as nominal, so each mismatch counts equally. We retain all nine judgments without aggregating by model family and weight instances equally, without rubric weights. Separate runs can share model-specific errors; these descriptive metrics do not establish statistical independence or judging accuracy.

\subsection{Human Agreement with Judges}
\label{app:human-agreement}

\textbf{Sample.} We select two tasks from each of the seven projects (14 tasks). For each task, we label the responses of two of three agents: GPT-6 Astra (A), Claude Opus 5.5 (B), and Gemini 3.8 Flash (C), all at high effort. Agent pairs are assigned at random, giving A+B and A+C five tasks each and B+C four. The 28 responses contain 852 response--criterion instances, 454 for gap identification and 398 for revision. Reviewers see agents only by letter.

\textbf{Labeling.} Each human reviewer independently labeled all 852 instances without access to the panel's verdicts or the other reviewer's labels. A reviewer receives the same inputs as a judge, the rubric and the anonymized prediction file, and assigns each criterion 1 (satisfied) or 0 (not satisfied). Reviewer 1 and Reviewer 2 mark 27.7\% and 32.3\% of criteria as satisfied, respectively, compared with 22.4\% for the panel.

\textbf{Metrics.} The panel label is the scoring consensus: satisfied when at least five of the nine judgments are SATISFIED. A judge model's label is the majority of its three trials. Cohen's $\kappa$~\citep{cohen1960kappa} corrects observed agreement for the agreement expected from each rater's own label frequencies. Criteria of one response share its wording and are not independent, so 95\% intervals come from 10,000 bootstrap resamples of whole responses. Task scores apply each task's tree weights (Appendix~\ref{app:score-definition}) to the binary labels.

\begin{table}[t]
    \centering\footnotesize
    \setlength{\tabcolsep}{2.5pt}
    \caption{Agreement on 852 sampled response--criterion instances, with 95\% intervals from resampling responses. R1 and R2 are the human reviewers; ``agreed'' restricts to the 763 instances where they agree.}
    \label{tab:human-agreement}
    \begin{tabular}{@{}lrll@{}}
        \toprule
        Comparison & $n$ & Agreement (\%) & Cohen's $\kappa$ \\
        \midrule
        R1 vs.\ R2 & 852 & 89.6 [87.4, 91.9] & 0.75 [0.70, 0.81] \\
        Single judge vs.\ R1, R2 & 852 & 91.3 / 88.5 & 0.77 / 0.72 \\
        Single judge vs.\ human & 852 & 89.9 & 0.74 \\
        Panel vs.\ R1 & 852 & 92.4 [90.3, 94.6] & 0.80 [0.75, 0.85] \\
        Panel vs.\ R2 & 852 & 89.7 [88.1, 91.1] & 0.74 [0.70, 0.78] \\
        Panel vs.\ agreed & 763 & 95.8 [94.3, 97.2] & 0.89 [0.85, 0.92] \\
        \addlinespace
        Judge model pairs & 852 & 91.3--93.4 & 0.76--0.82 \\
        \bottomrule
    \end{tabular}
\end{table}

\textbf{Observations.} The nine-judgment panel agrees with each human reviewer more than a single judgment does (Table~\ref{tab:human-agreement}), which supports aggregating judgments. Disagreements lean toward withholding credit: against Reviewer 1, 55 of the panel's 65 mismatches withhold credit; against Reviewer 2, 86 of 88 do; and against the reviewers' agreed label, all 32 do. Across the 28 responses, tree-weighted task scores under the panel correlate strongly with both reviewers (Pearson $r = 0.88$ and $0.87$), and when the two agents' scores are ordered within each task, the panel matches Reviewer 1's order on 12 of 14 tasks and Reviewer 2's on 11.

\section{Case Study: Kubernetes Gang Scheduling}
\label{app:gang-running-example}

This appendix follows the gang scheduling task (KEP-4671, Kubernetes enhancements PR 5558) through \projectname{}. Its rubric has 70 criteria, 36 deficiency and 34 revision criteria. We use the response of the highest-scoring agent, Claude Opus 5.5 (max), which submits 32 prediction pairs and scores 0.525 on this task.

\begin{figure}[t]
\centering
\includegraphics[width=\columnwidth]{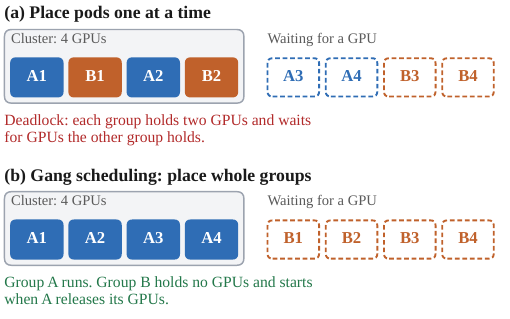}
\caption{Two pod groups each need four GPUs on a four-GPU cluster. (a) Placing pods one at a time lets each group hold part of the cluster, and neither can run. (b) Gang scheduling lets one group run while the other waits without holding GPUs.}
\label{fig:gang-scheduling-scenario}
\end{figure}

\begin{figure*}[t]
\begin{tcolorbox}[promptbox,unbreakable,title={KEP-4671 initial proposal (excerpts; ``[\ldots]'' marks omissions)}]
\textbf{Summary} \hfill\textcolor{gray}{\emph{group placement, waiting timeout}}\\
In this KEP, kube-scheduler is modified to support gang scheduling. To implement gang scheduling, kube-scheduler identifies pods that are in a group and waits until all pods reach the same stage of the scheduling/binding cycle before allowing any pods from the group to advance past that point. If not all pods can reach that point before a timeout expires, then the scheduler stops trying to schedule that group, and all pods release all their resources. This allows other workloads to try to allocate those resources.

\textbf{Goals} \hfill\textcolor{gray}{\emph{group placement}}
\begin{itemize}[nosep,leftmargin=*]
\item[{}] [\ldots]
\item When two workloads of the same priority, and both requiring gang scheduling are submitted, a deadlock will not occur.
\begin{itemize}[nosep,leftmargin=*]
\item In particular, avoid the case where where each workload holds some resources, and is waiting for the other to release enough resources for it to complete.
\end{itemize}
\end{itemize}

\textbf{Naming} \hfill\textcolor{gray}{\emph{group definition}}
\begin{itemize}[nosep,leftmargin=*]
\item[{}] [\ldots]
\item Within a Workload there is a list of groups of pods. Each group represents a top-level division of pods within a Workload. Each group can be independently gang scheduled (or not use gang scheduling). This group is named \texttt{<<[UNRESOLVED community feedback requested]>>} \texttt{PodGroup} or \texttt{GangGroup} for the top level. \texttt{<<[/UNRESOLVED]>>}.
\end{itemize}

\textbf{Proposal} \hfill\textcolor{gray}{\emph{waiting timeout}}\\
The \texttt{Workload} core resource will be introduced. A \texttt{Workload} does not create any pods. It just describes what pods the scheduler should expect to see, and how to treat them. [\ldots] A sample resource looks like this:

\noindent\begin{minipage}[t]{0.48\linewidth}
\vspace{0pt}
\begin{Verbatim}[fontsize=\footnotesize]
apiVersion: scheduling/v1alpha1
kind: Workload
metadata:
  name: myjob
spec:
  podGroups:   # or gangGroups -- TBD
    - name: "pg1"
      gangMode: Single
      minCount: 100
      schedulingTimeoutSeconds: 60
\end{Verbatim}
\end{minipage}\hfill
\begin{minipage}[t]{0.48\linewidth}
\vspace{0pt}
\textbf{API} \hfill\textcolor{gray}{\emph{waiting timeout}}
\begin{Verbatim}[fontsize=\footnotesize]
// GangSchedulingPolicy holds options
// that affect how gang scheduling of one
// PodGroup is handled by the scheduler.
type GangSchedulingPolicy struct {
    ShedulingTimeoutSeconds *int
    MinCount *int
}
\end{Verbatim}
\end{minipage}

\textbf{Scheduler Changes} \hfill\textcolor{gray}{\emph{group placement}}\\
The Gang Scheduling functionality will be implemented using a combination of kube-scheduler plugins and scheduling framework changes. The plugin will use PreEnqueue and WaitOnPermit framework hooks to control the scheduling process of a gang. The scheduling framework changes will watch Workloads, map pods to and from their Workloads and Gang(s) within the workload.

More detail about scheduled changes is described in this document. [\ldots]
\end{tcolorbox}
\caption{Excerpts of the initial gang scheduling proposal. Gray tags name the design decision of Section~\ref{sec:gang-example} that each excerpt states.}
\label{fig:gang-initial-rfc}
\end{figure*}

\begin{table*}[t]
\centering\footnotesize
\renewcommand{\arraystretch}{1.1}
\caption{Selected concerns from the expert review of KEP-4671, condensed.}
\label{tab:gang-review}
\begin{tabularx}{\textwidth}{@{}>{\raggedright\arraybackslash}p{2.3cm}>{\raggedright\arraybackslash}X>{\raggedright\arraybackslash}X@{}}
\toprule
Concern & Deficiency raised & Revision reached \\
\midrule
Group definition & A reviewer asks whether one group's failure or timeout should affect the other groups of the same application. & Groups are independent: a group that can be scheduled runs even if another cannot. Groups that need each other must form a single group. \\
Group placement & Placing pods one at a time cannot by itself prevent the deadlock that the proposal's goals rule out, and can miss feasible placements. & Alpha keeps the simple mechanism. Beta places all pods of a group together against one snapshot of the cluster, validates the placement through the normal scheduler phases, and recomputes the whole placement if any pod fails. \\
Waiting timeout & A timeout set by each application is hard for users to tune and lets one job deny service to others by holding resources. & Remove the field from the alpha API; any timeout belongs in the scheduler's cluster-level configuration. \\
Preemption & Preempting pods one at a time can evict running pods for a group that never completes. & Beta performs no preemption for a group until a feasible placement exists for every pod. Preempting a running group as a unit is a separate future feature. \\
Group membership & Pods can join a group through either label selectors on the application or a field on the pod, and the proposal leaves the choice open. & Each pod carries an explicit, immutable reference to its application and group. Size fields may change, but a change that races with a scheduling decision need not affect it. \\
\bottomrule
\end{tabularx}
\end{table*}

\subsection{Gang Scheduling Background}
\label{app:gang-background}

Figure~\ref{fig:gang-scheduling-scenario} shows the deadlock that gang scheduling prevents.

\begin{table*}[t]
\centering\footnotesize
\renewcommand{\arraystretch}{1.05}
\caption{Selected final rubric criteria for the gang scheduling task, verbatim.}
\label{tab:gang-rubric}
\begin{tabularx}{\textwidth}{@{}l>{\raggedright\arraybackslash}X@{}}
\toprule
Type & Rubric criterion \\
\midrule
\multicolumn{2}{@{}l}{\emph{Group definition}} \\
Deficiency & The agent must identify that the RFC allows multiple PodGroups in one Workload but does not define whether a schedulable group may proceed when another group is below minCount or unschedulable, leaving the all-or-nothing boundary ambiguous. \\
Revision & The agent must state that separate PodGroups, including replicated gang instances, schedule independently, so one schedulable group may run even when another group in the same Workload cannot. \\
\midrule
\multicolumn{2}{@{}l}{\emph{Group placement}} \\
Deficiency & The agent must identify that the RFC presents an alpha PreEnqueue/WaitOnPermit pod-by-pod gang scheduling mechanism without explaining how ordering, deadlock, and placement-quality limitations will be avoided. \\
Revision & The agent must propose that beta add a workload-scheduling phase that considers every pod in a gang against one cluster-state snapshot and computes their placement together before ordinary pod-by-pod scheduling. \\
Revision & The agent must specify that if any gang pod cannot validate or execute its proposed placement, the whole attempt is rejected and placement is recomputed for every gang pod, rather than retaining a partial gang in WaitOnPermit. \\
\midrule
\multicolumn{2}{@{}l}{\emph{Waiting timeout}} \\
Deficiency & The agent must identify that public per-PodGroup schedulingTimeoutSeconds lacks a clear start point and expiration behavior. \\
Deficiency & The agent must identify that public per-PodGroup schedulingTimeoutSeconds is difficult for users to tune and can allow jobs to monopolize held resources. \\
Revision & The agent must propose removing schedulingTimeoutSeconds from the alpha Workload API and keeping any necessary timeout in kube-scheduler configuration or an internal default until a clear per-workload use case exists. \\
\midrule
\multicolumn{2}{@{}l}{\emph{Preemption}} \\
Deficiency & The agent must identify that ordinary pod-by-pod preemption could evict or reserve capacity for only part of a high-priority gang before all members are feasible, wasting capacity and disrupting running workloads if the gang never completes. \\
Revision & The agent must specify that beta performs no physical preemption for a gang until a feasible placement for every member is computed and validated; if preemption cannot make one member feasible, none of the gang's preemptions execute. \\
\midrule
\multicolumn{2}{@{}l}{\emph{Group membership and updates}} \\
Deficiency & The agent must identify that the RFC leaves pod-to-PodGroup association unresolved between Workload-side label selectors and an explicit Pod field, creating ambiguous membership and mismatch handling. \\
Revision & The agent must propose that each participating Pod carry an explicit WorkloadReference in PodSpec containing the Workload and PodGroup names, rather than inferring membership through Workload-side selectors. \\
Revision & The agent must state that sizing and similar non-reference Workload fields may change over time, while an update racing an in-flight decision is not guaranteed to affect that decision. \\
\midrule
\multicolumn{2}{@{}l}{\emph{Comparison with existing tools (unresolved)}} \\
Deficiency & The agent must identify that, although the RFC claims native support would provide a stock-cluster baseline and standard interface, it does not provide the requested concrete comparison explaining what core kube-scheduler gang scheduling adds over existing tools such as Kueue or Volcano. \\
\bottomrule
\end{tabularx}
\end{table*}

\subsection{Initial Proposal}
\label{app:gang-initial-rfc}

Figure~\ref{fig:gang-initial-rfc} reproduces the parts of the initial proposal that state the three design decisions of Section~\ref{sec:gang-example}, with the proposal's original wording and typos.

\begin{table*}[t]
\centering\footnotesize
\renewcommand{\arraystretch}{1.1}
\caption{Five of the 32 prediction pairs from Claude Opus 5.5 (max), condensed, with the criteria the judges matched to each. Counts give the verdicts, out of nine, that support each outcome.}
\label{tab:gang-predictions}
\begin{tabularx}{\textwidth}{@{}r>{\raggedright\arraybackslash}X>{\raggedright\arraybackslash}X>{\raggedright\arraybackslash}p{4.9cm}@{}}
\toprule
\# & Predicted deficiency & Predicted revision & Matched criteria \\
\midrule
9 & The goals promise that two gangs of the same priority will not deadlock, but holding pods at the Permit stage lets two gangs each hold part of the capacity until the timeout and then collide again; placing pods greedily can also miss a feasible placement. & State only what alpha guarantees (no gang is partially bound; reservations are released after the timeout), document the livelock and placement risks, and plan a later scheduling cycle that evaluates all pods of a group together. & \sat\ Pod-by-pod scheduling limitations (8/9)\newline \sat\ Alpha-to-beta architecture gap (6/9)\newline \miss\ Gang-wide beta placement (1/9) \\
6 & The timeout is contradictory and underspecified: when the timer starts, what happens after expiry, and its default and bounds are undefined, and ``stops trying'' contradicts the goal that gangs never remain stuck. & Remove \texttt{schedulingTimeoutSeconds} from the alpha API and use a fixed plugin timeout; on expiry, reject all waiting pods, release their resources, and requeue them with backoff. & \sat\ Remove public alpha timeout (9/9)\newline \sat\ Undefined timeout lifecycle (9/9)\newline \sat\ Continuous timeout retries (9/9)\newline \miss\ Timeout tuning and monopolization (0/9) \\
1 & The mapping from pods to groups is unresolved, and the label-selector design lets pods silently move between groups and lets selectors overlap. & Give each pod a reference to its Workload, its group, and an optional replica key; pods with different replica keys form independent gangs. Remove the selectors. & \sat\ Ambiguous group association (9/9)\newline \sat\ Explicit WorkloadReference membership (9/9)\newline \miss\ Independent group scheduling (3/9) \\
8 & The proposal does not say what happens when a referenced Workload is missing, deleted, or updated, or who creates Workloads. & Hold pods until their Workload appears; make the group list and its policies immutable in alpha; let users or external controllers create Workloads. & \sat\ Missing Workload behavior gap (9/9)\newline \sat\ Hold until Workload observation (9/9)\newline \con\ Mutable sizing with race limits (9/9) \\
17 & The proposal ignores preemption: a gang member can evict other pods although its gang may later time out, wasting the disruption. & Keep pod-by-pod preemption in alpha and defer preemption that evicts only when the whole gang fits to a follow-up proposal. & \sat\ Partial preemption risk (9/9)\newline \miss\ All-or-nothing gang preemption (contradicted 3/9) \\
\bottomrule
\end{tabularx}
\end{table*}

\subsection{Expert Review}
\label{app:gang-review}

Table~\ref{tab:gang-review} summarizes how the experts resolved five concerns. The review separated the first (alpha) release from the next (beta): it kept the simple pod-by-pod mechanism for alpha and required whole-group placement for beta.

\subsection{Reference Rubric}
\label{app:gang-rubric}

Table~\ref{tab:gang-rubric} reproduces 15 of the 70 criteria. The last one is unresolved: the experts asked for a comparison with existing tools but never settled on one, so it has no revision criterion.

Figure~\ref{fig:gang-tree} shows the weighted tree, scored for the example agent. It expands the two branches that contain the three decisions (highlighted) and collapses most other subtrees to their counts.

\begin{figure*}[p]
\centering
\begin{minipage}[t]{0.49\textwidth}
\vspace{0pt}
\resizebox{\linewidth}{!}{\begin{forest} gangtree
[{\nd{}{Complete specification analysis}{0.525}}
  [{\nd{3}{Gang scheduling architecture}{0.380}}
    [{\nd{3}{Scheduling architecture evolution}{0.313}\ \dec{group placement}}, hl
      [{\lf{\sat}{D}{2}{Alpha-to-beta architecture gap}}]
      [{\lf{\sat}{D}{3}{Pod-by-pod scheduling limitations}}]
      [{\lf{\miss}{R}{2}{Alpha scope and intent}}]
      [{\lf{\miss}{R}{3}{Gang-wide beta placement}}]
      [{\lf{\miss}{R}{3}{Normal-phase placement validation}}]
      [{\lf{\miss}{R}{3}{Whole-gang recomputation}}]
    ]
    [{\nd{2}{Independent PodGroup scheduling}{0.000}\ \dec{group definition}}, hl
      [{\lf{\miss}{D}{2}{Ambiguous all-or-nothing boundary}}]
      [{\lf{\miss}{R}{3}{Independent group scheduling}}]
    ]
    [{\nd{2}{Timeout retry semantics}{1.000}}
      [{\lf{\sat}{D}{2}{Undefined post-timeout outcome}}]
      [{\lf{\sat}{R}{3}{Continuous timeout retries}}]
    ]
    [{\nd{3}{Atomic gang preemption}{0.400}}
      [{\lf{\sat}{D}{2}{Partial preemption risk}}]
      [{\lf{\miss}{R}{3}{All-or-nothing gang preemption}}]
    ]
    [{\nd{2}{Gang-aware preemption scope}{0.400}}
      [{\lf{\miss}{D}{3}{Undefined effective gang priority}}]
      [{\lf{\sat}{D}{2}{Undefined victim-gang atomicity}}]
      [{\lf{\sat}{R}{2}{Future victim-gang atomicity}}]
      [{\lf{\miss}{R}{3}{Beta preemption design milestone}}]
    ]
    [{\lf{\miss}{D}{1}{Stalled preemption recomputation gap}}]
  ]
  [{\nd{3}{Workload API and membership model}{0.561}}
    [{\cnd{3}{Explicit PodGroup membership}{1.000}{2}{2}}]
    [{\cnd{2}{Replicated group identity}{1.000}{2}{2}}]
    [{\cnd{2}{Namespace and reference scope}{1.000}{2}{2}}]
    [{\cnd{1}{Grouping type names}{0.500}{1}{2}}]
    [{\cnd{3}{Scheduling policy API shape}{1.000}{2}{2}}]
    [{\nd{2}{API mutability and races}{0.357}}
      [{\lf{\sat}{D}{2}{Pod reference mutability gap}}]
      [{\lf{\sat}{D}{2}{WorkloadSpec mutability gap}}]
      [{\lf{\sat}{D}{2}{Concurrent-update semantics gap}}]
      [{\lf{\miss}{R}{3}{Immutable Pod WorkloadReference}}]
      [{\lf{\sat}{R}{2}{Immutable ControllerRef}}]
      [{\lf{\con}{R}{3}{Mutable sizing with race limits}}]
    ]
    [{\cnd{2}{PodGroup member diversity}{0.000}{0}{3}}]
    [{\cnd{3}{General Workload semantics}{0.000}{0}{2}}]
    [{\cnd{2}{Controller provenance}{0.000}{0}{2}}]
  ]
]
\end{forest}}
\end{minipage}%
\hfill\begin{minipage}[t]{0.49\textwidth}
\vspace{0pt}
\resizebox{\linewidth}{!}{\begin{forest} gangtree
[{\nd{}{Complete specification analysis (continued)}{}}
  [{\nd{2}{Lifecycle and scheduler integration}{0.778}}
    [{\nd{2}{Missing Workload handling}{1.000}}
      [{\lf{\sat}{D}{2}{Missing Workload behavior gap}}]
      [{\lf{\sat}{R}{3}{Hold until Workload observation}}]
    ]
    [{\nd{3}{Workload lifecycle ownership}{0.778}}
      [{\lf{\sat}{D}{2}{Unassigned lifecycle responsibility}}]
      [{\lf{\sat}{D}{2}{Unspecified controller propagation}}]
      [{\lf{\sat}{R}{3}{User-managed alpha lifecycle}}]
      [{\lf{\miss}{R}{2}{Nonduplicated Job references}}]
    ]
    [{\nd{2}{Scheduling timeout configuration}{0.556}\ \dec{waiting timeout}}, hl
      [{\lf{\sat}{D}{2}{Undefined timeout lifecycle}}]
      [{\lf{\miss}{D}{2}{Timeout tuning and monopolization}}]
      [{\lf{\miss}{D}{2}{Timeout and recomputation conflict}}]
      [{\lf{\sat}{R}{3}{Remove public alpha timeout}}]
    ]
    [{\nd{3}{Scheduler watches and gates}{0.778}}
      [{\lf{\miss}{D}{2}{Undefined Workload observation model}}]
      [{\lf{\sat}{D}{2}{Undefined scheduling-gate roles}}]
      [{\lf{\sat}{R}{2}{Direct Workload watches}}]
      [{\lf{\sat}{R}{3}{Distinct alpha scheduling gates}}]
    ]
  ]
  [{\nd{2}{Compatibility and operational readiness}{0.436}}
    [{\cnd{3}{Existing Pod constraints}{0.000}{0}{2}}]
    [{\cnd{2}{Cluster autoscaler scope}{0.500}{1}{2}}]
    [{\nd{2}{Status and observability}{0.400}}
      [{\lf{\sat}{D}{3}{Missing user-visible gang state}}]
      [{\lf{\sat}{D}{1}{Missing scheduler observability}}]
      [{\lf{\miss}{R}{2}{Workload status by beta}}]
      [{\lf{\miss}{R}{2}{Rejected-placement metric}}]
      [{\lf{\miss}{R}{2}{Whole-gang latency metric}}]
    ]
    [{\cnd{3}{Enablement and rollback safety}{1.000}{6}{6}}]
    [{\lf{\miss}{D}{1}{Missing ecosystem comparison}}]
  ]
]
\end{forest}}

\vspace{1.2em}
\begin{tcolorbox}[promptbox,unbreakable,title={Legend},width=0.92\linewidth]
\begin{tabular}{@{}ll@{}}
\sat & Criterion satisfied (score $+1$) \\
\con & Criterion contradicted (score $-1$) \\
\miss & Criterion not satisfied (score 0) \\
D / R & Deficiency / revision criterion \\
{[}$w${]} & Weight relative to siblings \\
\textcolor{gray}{0.556} & Node score for the example agent \\
\textcolor{gray}{$k$/$n$} & Collapsed subtree: $k$ of $n$ satisfied \\
\colorbox{hlcolor}{\strut\ \ } & Decision from Section~\ref{sec:gang-example} \\
\end{tabular}
\end{tcolorbox}
\end{minipage}
\caption{The weighted rubric tree of the gang scheduling task, scored for Claude Opus 5.5 (max). Leaves show short titles of the criteria; Table~\ref{tab:gang-rubric} gives the full text of selected criteria.}
\label{fig:gang-tree}
\end{figure*}

\subsection{Agent Predictions and Judgments}
\label{app:gang-predictions}

Table~\ref{tab:gang-predictions} condenses five of the agent's 32 prediction pairs and lists the criteria that the judges matched to each.

The three decisions of Section~\ref{sec:gang-example} have different outcomes. For \emph{group placement}, the agent finds the deficiency (prediction 9) but defers whole-group placement to an unspecified later release instead of proposing the experts' beta design. For the \emph{waiting timeout}, it proposes the experts' revision, removing the field, but because the field is underspecified, not because it lets one job hold resources that others need. For \emph{group definition}, it never asks whether one group may run while another group of the same application cannot.

The judges also find two contradictions. Prediction 8 makes the group list immutable, whereas the experts kept size fields mutable. All nine verdicts mark a contradiction, so the criterion scores $-1$. Prediction 17 defers the protection against partial preemption that the experts required for beta. Only three verdicts mark a contradiction, so the criterion scores 0.

\subsection{Score Computation}
\label{app:gang-score}

Each node's score in Figure~\ref{fig:gang-tree} is the weighted mean of its children's scores. The timeout node has four leaves with weights 2, 2, 2, and 3, of which the first and last are satisfied:
\[
s_{\mathrm{timeout}} = \frac{2\cdot 1 + 2\cdot 0 + 2\cdot 0 + 3\cdot 1}{2+2+2+3} = \frac{5}{9} = 0.556.
\]
Its parent, lifecycle and scheduler integration, has four children with weights 2, 3, 2, and 3:
\[
\frac{2\cdot 1.000 + 3\cdot 0.778 + 2\cdot 0.556 + 3\cdot 0.778}{10} = 0.778.
\]
The root combines its four branches with weights 3, 3, 2, and 2 into the task score:
\[
\frac{3\cdot 0.380 + 3\cdot 0.561 + 2\cdot 0.778 + 2\cdot 0.436}{10} = 0.525.
\]

The contradiction lowers the API mutability node from $8/14 = 0.571$ to $(8-3)/14 = 0.357$, and the task score from 0.531 to 0.525. Restricting the tree to deficiency or revision criteria gives an identification score of 0.616 and a revision score of 0.449.

\end{document}